\documentclass[lettersize,journal]{IEEEtran}
\usepackage{amsmath,amssymb,amsfonts}
\usepackage{algorithmic}
\usepackage{algorithm}
\usepackage{array}
\usepackage[caption=false,font=normalsize,labelfont=sf,textfont=sf]{subfig}
\usepackage{textcomp}
\usepackage{stfloats}
\usepackage{verbatim}
\usepackage{graphicx}
\usepackage{cite}

\usepackage{xcolor}

\usepackage{hyperref}
\usepackage{pifont}

\usepackage{caption}

\usepackage{units}
\usepackage{longtable,array}
\usepackage{lscape}
\usepackage{arydshln}
\usepackage{multicol}
\usepackage{multirow}

\usepackage{xurl}

\usepackage{breqn}
\usepackage{soul}

\newcolumntype{P}[1]{>{\centering\arraybackslash}p{#1}}

\newcommand{\xmark}{\ding{55}}%
\begin{document}

\title{Inertial Sensors for Human Motion Analysis: A Comprehensive Review}

\author{Sara García-de-Villa, David Casillas-Pérez, Ana Jiménez-Martín \IEEEmembership{Member, IEEE}, and Juan Jesús García-Domínguez \IEEEmembership{Senior Member, IEEE}

\thanks{November 2\textsuperscript{nd}, 2022. Financial support acknowledgment. This work was supported in part by Junta de Comunidades de Castilla La Mancha (FrailAlert project, SBPLY/21/180501/000216), the
Spanish Ministry of Science and Innovation (INDRI project, PID2021-122642OB-C41) , Comunidad de Madrid (RACC project, CM/JIN/2021-016) and NEXTPERCEPTION European Union project funded by ECSEL Joint Undertaking (JU), under grant agreement No.876487 (ECSEL-2019-2-RIA). }
\thanks{S. G. V. was with the Department of Electronics, University of Alcala, Alcalá de Henares, 28805, Spain and in the Center of Automation and Robotics (CSIC-UPM), Arganda del Rey, 28500, Spain  (e-mail: sara.garciavilla@uah.es).  She is now with the Department of Signal Theory and Communications, Rey Juan Carlos University, Fuenlabrada, 28942, Spain.  (e-mail: sara.garcia.devilla@urjc.es). }
\thanks{A. J. M. and J. J. G. D. are with the Department of Electronics, University of Alcala, Alcalá de Henares, 28805, Spain.  (A. J. M. e-mail: ana.jimenez@uah.es; J. J. G. D. e-mail: jjesus.garcia@uah.es).}
\thanks{D. C. P. is with the Department of Signal Theory and Communications, Rey Juan Carlos University, Fuenlabrada, 28942, Spain.  (e-mail: david.casillas@urjc.es).}}

\markboth{IEEE Transactions on Instrumentation and Measurement,~Vol.~X, No.~X, November~2022}%
{García-de-Villa \MakeLowercase{\textit{et al.}}: Inertial Sensors for Human Motion Analysis: A Comprehensive Review}

\maketitle

\begin{abstract}
Inertial motion analysis is having a growing interest during the last decades due to its advantages over classical optical systems.  The technological solution based on inertial measurement units allows the measurement of movements in daily living environments, such as in everyday life, which is key for a realistic assessment and understanding of movements. This is why research in this field is still developing and different approaches are proposed. This presents a systematic review of the different proposals for inertial motion analysis found in the literature. The search strategy has been carried out in eight different platforms, including journal articles and conference proceedings, which are written in English and published until August 2022. The results are analyzed in terms of the publishers, the sensors used, the applications, the monitored units, the algorithms of use, the participants of the studies and the validation systems employed. In addition, we delve deeply into the machine learning techniques proposed in recent years and in the approaches to reduce the estimation error. In this way, we show an overview of the research carried out in this field, going into more detail in recent years, and providing some research directions for future work. 
\end{abstract}

\begin{IEEEkeywords}
Human motion, motion analysis, kinematic analysis, inertial measurement units, IMU, inertial sensors.
\end{IEEEkeywords}

\section{Introduction}
\label{sec:introduction}

Human motion analysis is an essential support tool for the assessment of the
parameters of movements, which is specially important in the evaluation of workout routines, clinical rehabilitation and preventive treatments~\cite{Fitzgerald2007}. 
It is also becoming very popular for the physical activity monitoring in the elderly. Indeed, as the population of 
developed countries ages, the demand for home-based rehabilitation and the need to obtain quantitative exercise data remotely will increase~\cite{1501143}. 

Optical methods are considered the \emph{gold standard} in the motion analysis field because of their accurate measurements of kinematic and spatio-temporal parameters~\cite{camomilla2018trends}. However, these systems entail several disadvantages, such as the high cost of equipment, the need for trained personnel to use the equipment, the required large spaces for installation and their restricted margin of maneuverability, that limits their use to controlled indoor environments.

The inertial motion analysis has emerged as a promising alternative to optical methods attracting a great scientific interest.
Inertial systems are portable and can be used everywhere, what means an advantage to the optical systems, which are commonly constraint to a limited space.
That makes the Inertial Measurement Units (IMUs)  an affordable and friendly use alternative for the estimation of human kinematics. 
These devices allow continuous monitoring of human motions in daily environments, which is crucial in order to obtain more reliable information than the obtained in sporadic laboratory tests. 
For these reasons, the use of IMUs has increased in the last few decades for continuous monitoring of human motions, as reported in~\cite{lopez_nava_review}.

\begin{figure}[htbp]
    \centering
    \includegraphics[width=0.75\columnwidth]{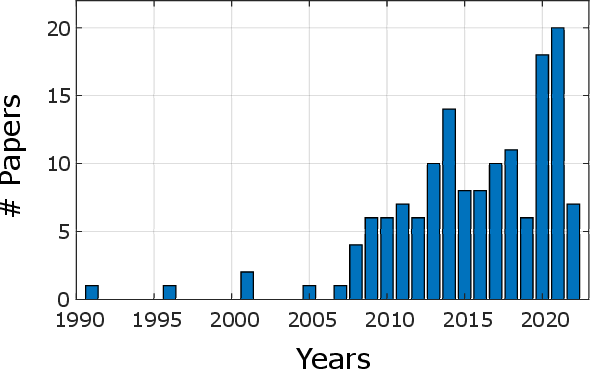}
    \caption{Number of publications focused on the inertial motion analysis, referred to obtaining kinematic parameters by using portable inertial sensors, found in the literature. 
    }
    \label{fig:interval_years}
\end{figure}

Previous works extensively review the use of portable sensors.
A recent review analyzes the integration of portable sensors in clothes to obtain physiological and motion information~\cite{kubicek2020_textilesclothigs}. 
However, inertial sensors are not considered in the analysis, in spite of their frequent use in this field. 
Reviews that take into account the use of inertial sensors are focused on applications, such as sign languages or motion analysis~\cite{kudrinko2020_signlanguage, camomilla2018trends}.
Works about the inertial motion analysis, as~\cite{camomilla2018trends}, address the motion monitoring  and kinematic feature extraction, but only considering the specific area of sport-related exercises evaluation and their analysis is up to April $2017$.
A lower-limb focused study is carried out in~\cite{weygers2020_review}, but it does not provide a complete overview of the literature on inertial motion analysis.
For the best of our knowledge, the last in-depth and generic systematic review on inertial sensors for human motion analysis is reported in~\cite{lopez_nava_review}, published in $2016$.
Since the number of publications about human motion analysis increases over time, as shown in Fig.~\ref{fig:interval_years}, we consider there is a need to update the literature review on this topic. 
According to Fig.~\ref{fig:interval_years}, the number of existing publications on the inertial motion analysis field 
has considerably increased since the previous review was published~\cite{lopez_nava_review}.

Furthermore, during the last years, Machine Learning (ML) methods have arisen and they have been applied to the inertial motion analysis. 
Consequently, it is required an update to provide an overview of the algorithms analyzed in~\cite{lopez_nava_review}, as the Kalman filters, complementary filters, integration, vector observation and others, but in combination with the novel ML-based approaches.

For these reasons, the main aim of this work is to review the current state of inertial sensors for human monitoring, especially considering the occurrence and evolution of ML methods for this research field. Another objective of this work is to analyze the current trends  and provide insights into inertial motion analysis. To do so, we review the published works on human motion analysis using IMUs and analyze the selected ones in terms of: \emph{1)} publisher and years, \emph{2)} sensors used, \emph{3)} type of estimations referred to the dimensions of the estimated magnitudes, \emph{4)} the aimed applications of the proposals, \emph{5)} the monitored motion units, \emph{6)} the algorithmic approaches, with an in-depth analysis of sensor fusion filters, data science algorithms and the approaches for error reduction, \emph{7)} the study participants and \emph{8)} the validation systems and metrics.
Finally, on the basis of the findings, we suggest future research directions.

The rest of this document is structured as follows:
Section~\ref{sec:methods} describes the search strategy and the eligibility criteria applied in this work;
Section~\ref{sec:review_findings} details and analyzes the findings according to the terms explained above;
Section~\ref{sec:discussion} discusses the general trends of the studied works and analyzes the future directions; and finally,
Section~\ref{sec:conclusions} summarizes the main contributions of this work.

\section{Materials and Methods}
\label{sec:methods}

In this section, we describe the workflow to search and select the works in the state-of-the-art included in this review.
We describe the paper screening process and analyze the common publishers of this field.
Finally, we detail the data  extracted from them for the further analysis.

\subsection{Eligibility Criteria}
\label{sub_methods:eligibility}

This review focuses on peer-reviewed articles, book chapters and conference papers.
Papers are required to be published in English and describe the methodology employed to obtain human kinematic parameters using only IMUs. Sensor fusion with other devices is not considered. This review only includes those papers that validate their results using a reference system. 
If a journal paper is an extended version of a conference one, only the journal paper is included.

\subsection{Literature Search Strategy}
\label{sub_methods:search_strategy}

Considering the eligibility criteria, we select eight databases (ACM Digital Library, IEEE Xplore, PubMed, Science Direct, Scopus, Taylor \& Francis Online, Web of Science and Wiley Online Library) for the search of related papers, see Table~\ref{tab:sources}. Following the strategy of the previous review~\cite{lopez_nava_review} about this topic, we use the same search command, which consists in:
\begin{center}
    \emph{(``human motion'' OR ``human movement'') AND (``wearable sensors'' OR ``inertial sensors'' OR ``wearable system'')},
\end{center}
considering their presence in the title, abstract or keywords.
The search includes journals, book chapters or conference proceedings. 
The paper abstract is required to be available during this search.
No restriction was imposed on the date of publications.

\begin{table}[ht]
\caption{Databases consulted in the literature search}
\label{tab:sources}
\begin{tabular}{llr}
\hline
Database                 & Source                         & \multicolumn{1}{l}{\#~papers} \\ \hline
ACM Digital Library      & dl.acm.org                     & $17$                                 \\ 
IEEE Xplore              & ieeexplore.ieee.org            & $145$                                \\ 
PubMed                   &  www.ncbi.nlm.nih.gov/pubmed    & $188$                               \\ 
ScienceDirect            & www.sciencedirect.com          & $115$                                 \\ 
Scopus                   & www.scopus.com/                & $1434$                               \\ 
Taylor \& Francis Online & www.tandfonline.com            & $2$                                  \\ 
Web of Science           & webofknowledge.com             & $326$                                \\ 
Wiley Online Library     & onlinelibrary.wiley.com        & $21$                                 \\ \hline
\end{tabular}
\end{table}

The initial search on the databases in  Table~\ref{tab:sources} leads to a review of \unit[$2\,248$]{papers}. The papers found in this search do not include important references from the state-of-the-art, as~\cite{Joukov2018} or~\cite{Adamowicz2019}, so we expand the search. The new search is only performed in the Scopus website, since it is the largest database of all those evaluated (see Table~\ref{tab:sources}). In this case, we use the following command, which is less restrictive than the previous one:
\begin{center}
\emph{((``human motion'' OR ``human movement'' OR ``joint kinematics'' OR ``body tracking'' OR kinematic* OR ``joint angle*'' OR ``joint angle velocity'' OR ``joint angle acceleration'') AND (imu OR ``inertial sensors'' OR ``inertial measurement unit'' OR accelerometer OR gyroscope OR magnetometer ))}.
\end{center} 
The search criteria is to find these key phrases in the title, abstract or keywords of articles. This second search adds $1\,882$ documents, so we finally obtain \unit[$4\,130$]{papers} to review.

\begin{figure}[htb!]
    \centering
    \includegraphics[width=\columnwidth]{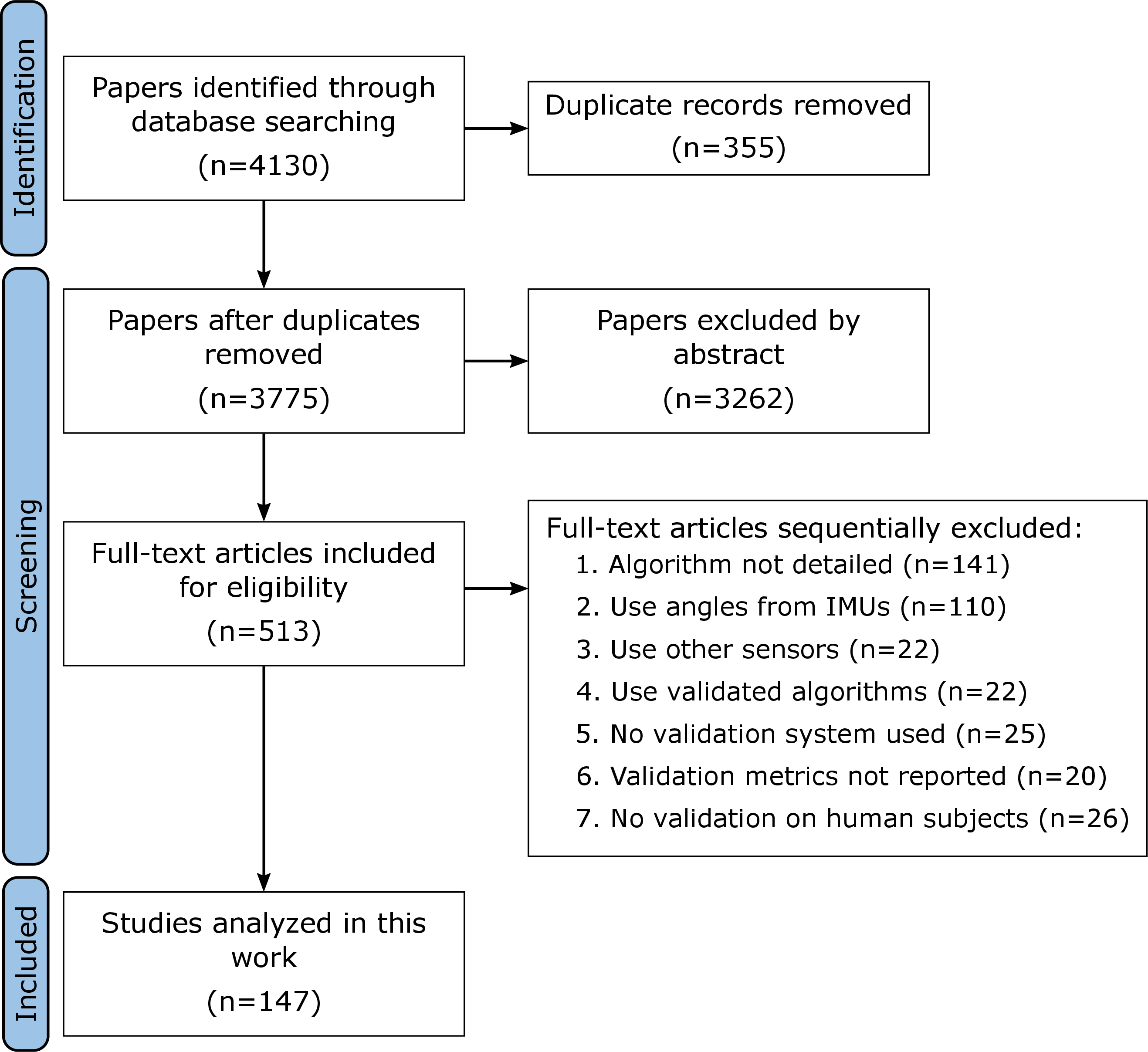}
    \caption{PRISMA search strategy flowchart. }
    \label{fig:papers_discrimination}
\end{figure}

Starting from the results of this search, we carry out a Preferred Reporting Items for Systematic reviews and Meta-Analyses (PRISMA) screening process~\cite{Page2021} to determine the documents included in this study. Fig.~\ref{fig:papers_discrimination} depicts the processes of identification and screening to determine the works included in this review.

\begin{figure*}[htbp]
    \centering
    \includegraphics[width=.7\textwidth]{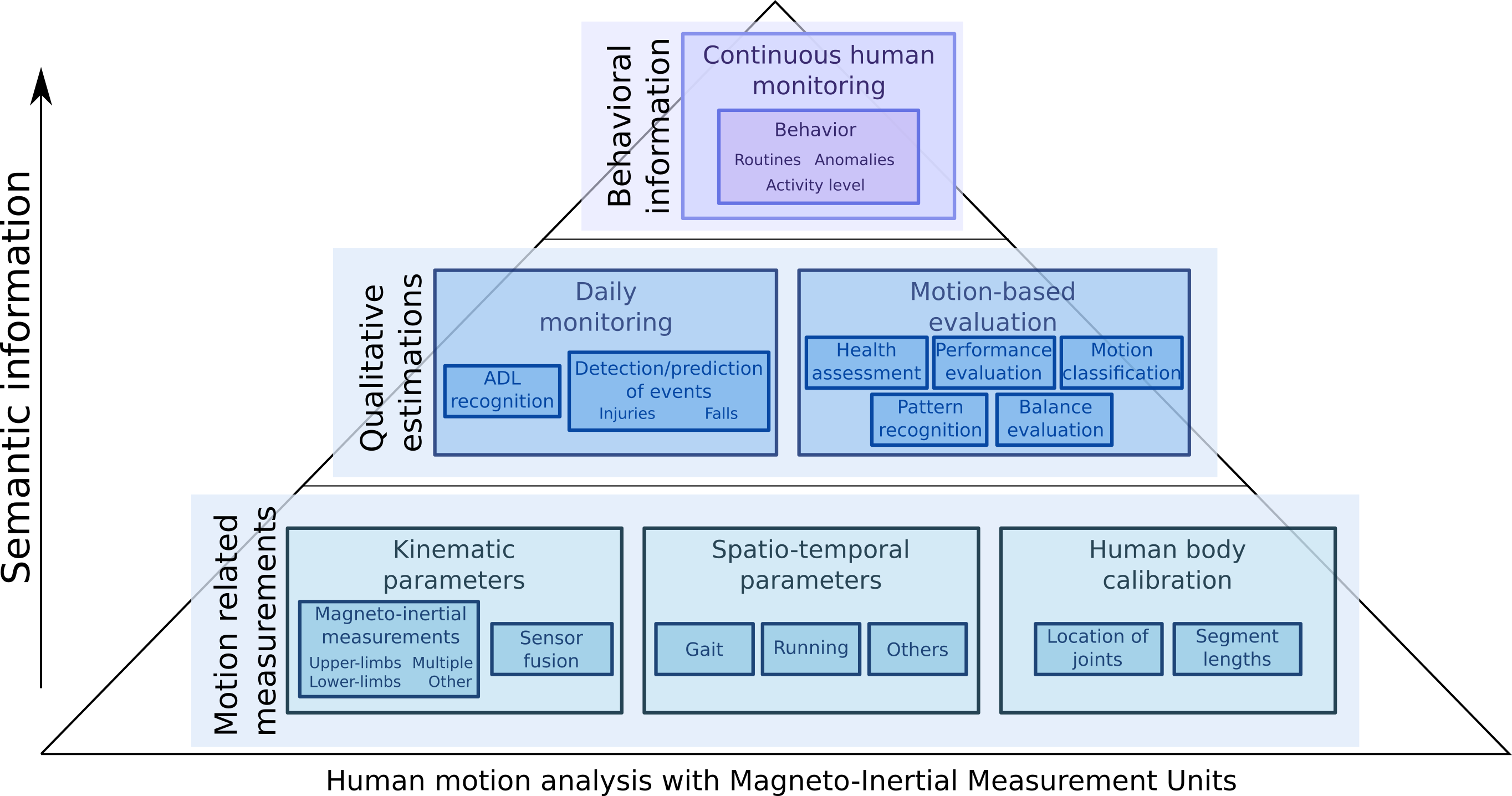}
    \caption{Distribution of the publications related to human motion analysis and IMUs, sorted by the semantic information obtained. This work focuses on the topic included in the left square of the lower row: kinematic parameters with magneto-inertial measurements.}
    \label{fig:papers_topics}
\end{figure*}

After excluding duplicated citations, the number of documents to screen is reduced to \unit[$3\,775$]{papers}. 
The results of this search include works in the field of human motion analysis with IMUs. 
However, this IMU-based motion analysis covers a wide range of topics, such as the estimation of kinematic and spatio-temporal parameters or the motion-based evaluation of health, as depicted in Fig.~\ref{fig:papers_topics}.

The topics of kinematic and spatio-temporal parameters refer to the analysis of different motion magnitudes, such as joint angles, trajectory or speed; whereas the human body calibration includes the location of joints or the estimation of segment lengths.
The last two topics, human monitoring and motion-based evaluation, are focused on qualitative analysis, such as recognizing types of motions or activities and identifying behavior patterns.
Our study is focused on wearable inertial sensors and kinematic parameters as joint rotation angles, so we discard by abstract reading those works that are focused on any other topic.
In this way, we exclude 
 \unit[$3\,262$]{papers} for not being related with the topic of this review.

We found \unit[$513$]{potentially} relevant studies of this topic for quality assessment. 
To consider a study in this review, we set the inclusion criteria detailed in Fig.~\ref{fig:papers_discrimination}, 
 which are referred to the proposed or applied algorithm, its validation and the sensor system used.  
Finally, \unit[$147$]{studies} meet the inclusion criteria and are analyzed in this review.

\subsection{Publisher and Years}
\label{sub_methods:publisherYears}

\begin{figure}[htbp!]
    \centering
    \includegraphics[width=0.77\columnwidth]{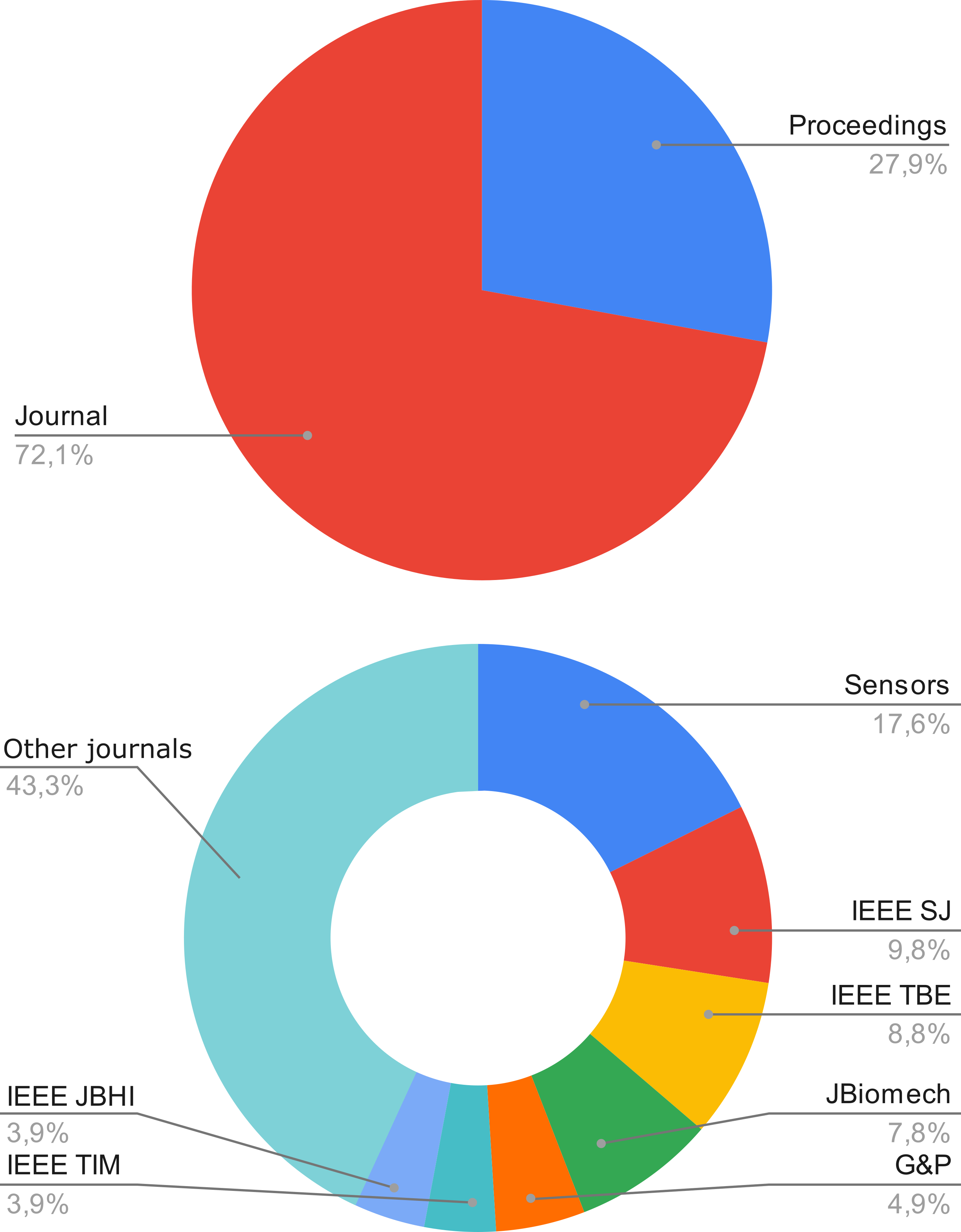}
    \caption{Distribution of the papers with respect to the type of publication environment in which they were published. Top: conference and journals distribution. Bottom: journals that published the analyzed works.}
    \label{fig:proceedings_journals}
\end{figure}

Most reviewed works are journal papers (\unit[$72.1$]{\%}) (see Fig.~\ref{fig:proceedings_journals}-top).
These works are published in \unit[$42$]{journals}.
The \unit[$56.7$]{\%} of them appear in \unit[$7$]{journals} (each of them with at least four papers), 
as shown in Fig.~\ref{fig:proceedings_journals}-bottom. The journals that appear with the highest frequency in this search are \emph{Sensors}, \emph{IEEE Sensors Journal} (IEEE SJ), \emph{IEEE Transactions on Biomedical Engineering} (IEEE TBE), \emph{Journal of Biomechanics} (JBiomech), \emph{Gait \& Posture} (G\&P), \emph{IEEE Transactions on Instrumentation and Measurement} (IEEE TIM) and \emph{IEEE Journal of Biomedical and Health Informatics} (IEEE JBHI). The remaining \unit[$43.3$]{\%} of works are distributed in \unit[$35$]{journals}.

There is a clear increasing interest in the research field of inertial motion capture (see Fig.~\ref{fig:interval_years}). 
This review is not restricted to any date in order to analyze all the works related to this topic and provide a general overview and its evolution. 
Fig.~\ref{fig:interval_years} shows the number of papers published during the 5-year periods from $1991$ until August $2022$. 
Only \unit[$2$]{works} are dated on the first studied decade, \unit[$21$]{works} on the second one and \unit[$98$]{works}  on the third one. 
There are also \unit[$27$]{works} published in the period $2021$-$2022$, the last years studied in this review.

\begin{figure}[ht!]
    \centering
    \includegraphics[width=0.8\columnwidth]{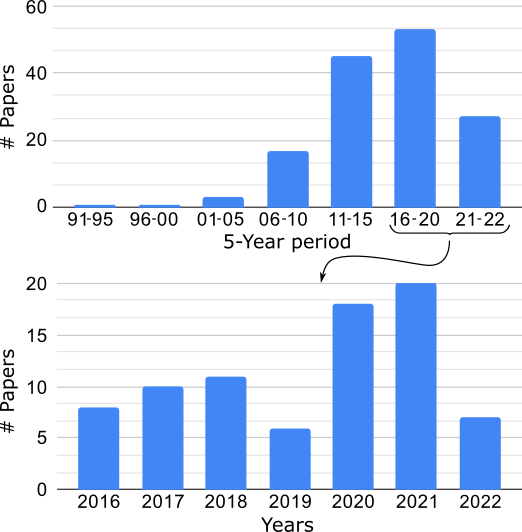}
    \caption{Year of publication of the reviewed papers. Top: trend since the first motion analysis related work until nowadays. 
    Bottom: distribution of publications in last 5-year period.}
    \label{fig:years}
\end{figure}

Since  $2016$, the last year studied in~\cite{lopez_nava_review}, the number of publications has highly (see Fig.~\ref{fig:years}).
These figures support the need for an update of a systematic review on the research topic of human motion analysis by using IMUs.

\subsection{Data Extraction, Analysis and Examination}
\label{sub_methods:data_extraction+analysis+examination}

We categorize the selected papers in terms of a set of relevant details. We classify them into two groups in order to ease the study and its reading. Firstly, we evaluate the details related with the implemented algorithms, the sensors in use and the estimations. And secondly, we study the specific anatomic part of the human body studied in each work, the validation system and metrics used, and the information related to the validation subjects.

Regarding the first set of details, we analyze the following parameters: the fusion algorithm (\textbf{FA}) implemented for the motion analysis, that indicates with ``SF'' if the work uses sensor fusion approaches, as ``ML'' the application of ML techniques and with ``OA'' any other proposal; 
the use of biomechanical constraints (\textbf{BC}) and their related requirements of anatomical information (\textbf{ANT}), as the segment lengths or the joint location with respect to the IMU sensors; 
the implementation of other corrections (\textbf{OC});
the type of sensor used to measure the motions (\textbf{GS}: gyroscope sensor, \textbf{AS}: accelerometer sensor, \textbf{MS}: magnetometer sensor) and the use of external sensors to train ML-based algorithms, but not in the motion prediction (\textbf{OS});
the type of estimation (\textbf{EST}), considering the possible planar (2D) or three-dimensional (3D) estimations;
the measured magnitude (\textbf{ANG}: angle, or \textbf{DIS}: displacement referred to the change in the position of the corresponding point, i.e. the sensor or the monitoring joint)
and the monitored motion unit (\textbf{JNT}: joint, or \textbf{SGM}: segment).
These details are shown in Table~\ref{tab:alg+sens+est} in Appendix~\ref{sec:appendix} for the selected papers.

With respect to the the human body part, we study the lower-group (\textbf{LG}) or upper-group (\textbf{UG}) of segments and joints. We also report the validation system (\textbf{VS}) used as \emph{ground truth} in the studied works and the metrics (RMSE: root mean square error;
nRMSE: normalized RMSE;
\%RMSE: percentage of RMSE;
MAE: mean absolute error; 
AE: average error;
CFC: correlation coefficient;
LAM: limits of agreement;
MV: maximum variation;
Accuracy; and
Error rate), labeled as \textbf{M1} and \textbf{M2} in Table~\ref{tab:part+validation+subjects}  in Appendix~\ref{sec:appendix},
employed in the proposed methods.
Finally, we provide the number of subjects (\textbf{NS}) studied and if this  population considered presents a motor-related disease (\textbf{DSS}).
Table~\ref{tab:part+validation+subjects} includes the details of these parameters above explained in the selected papers.

\section{Review findings}
\label{sec:review_findings}

Based on the categorization of the papers with respect to their relevant characteristics, presented in Table~\ref{tab:alg+sens+est} and Table~\ref{tab:part+validation+subjects}, in this section we describe the main findings.

\subsection{Sensors}
\label{sub_find:sensors}

\begin{figure}[htbp!]
    \centering
    \includegraphics[width=0.73\columnwidth]{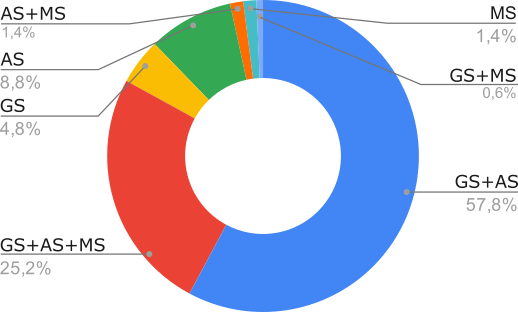}
    \caption{Sensor type and combination used in the analyzed works (AS: accelerometer sensor; GS: gyroscope sensor; MS: magnetometer sensor). 
    }
    \label{fig:sensor}
\end{figure}

IMUs contain tri-axial gyroscopes, accelerometers and, commonly, magnetometers. 
The information from these sensors are used separately through the observation of vectors, as gravity in the accelerometer data or the magnetic field in magnetometers, or by integration of the gyroscope data. 
Another approach is to gather their measurements in different combinations of two or three sensors with different algorithms, as sensor fusion filters or ML methods. 
In order to illustrate the proportion of their utilization, separately or fused, Fig.~\ref{fig:sensor} shows the percentage of use of each sensor or combination.

The integration of the turn rate alone entails inherent errors.
In the estimations of kinematic parameters, the turn rate integration results in an accumulated error  from the gyroscope bias.
For that reason, only \unit[$4.8$]{\%} of studies use this sensor alone~\cite{Allseits2018,Niijima2014,Muller2017,Grimpampi2013,An2012,Mohammadzadeh2015,Watanabe2016}.

Accelerometers are more frequently used separately (\unit[$8.8$]{\%}).
Their measurement of specific force allows us to obtain a direct observation of the gravity vector, used as orientation reference~\cite{Sun2016,Bagala2012,Laidig2017,Gholami2020,Das2014,Ma2015,Willemsen1991,Djuric-Jovicic2012,Liu2009c,Lim2020,Zehr2021,Zheng2021, Lee2010}. 
However, the direct observation of the gravity vector is only possible when accelerometers are static, as in gait strides during the stance phase.

Magnetometers are the most limited sensor analyzed because of their sensitivity to magnetic disturbances in the environment. As a consequence, only a \unit[$1.4$]{\%} of studies use this sensor independently~\cite{Watson20211,Friedman2014}. 

Sensor fusion techniques are useful methods to overcome the individual limitations of each of them separately.
Most of studies that fuse data from various sensors combine gyroscopes and accelerometers~\cite{Teruyama2013,Mundt2021,Ohtaki2001,Molnar2018,Dejnabadi2005,Allen2017,Yin2021,Ligorio2015,Choi2018,watanabe2011,Liu2020,Ligorio2020,Feldhege2015,Liu2009,Karunarathne2014,Hu2014,Goulermas2008,Mazza2012,Flores-Morales2016,Watanabe2014,Mundt20201,Tham2021,Joukov2019,Dorschky2020,Watanabe2015,Eom2014,Charry2011,Williamson2001,Bakhshi2011,Liu2009,Chen2020,Falbriard2020,Weygers2020,Meng2019,Joukov2020,Dorschky2019,Mundt2020,Rapp2021,Young2010,Lee2021,Jakob2013,Takeda2009,Xu2018,Cehajic2015,Kitano2019,Music2008,Lin2013,Lin2012,Seel2014,Baten1996,Cooper2009,Fasel2018,Bennett2013,Mayorca-Torres2020,Hernandez2021,Bonnet2016,Joukov2014,Liu2010,Findlow2008,Mundt2020b,Caroselli2013,Bonnet2013a,Villeneuve2017,Zhou2010,Joukov2015,Joukov2018,Alvarado2017,Alizadegan2017,ElGohary2012,Liang2021,Sharma2017,Ding2020,SharifiRenani2021,Tadano2013,Kumar2018,ElGohary2011,Zhou2008,Salehi2020,Figueiredo2020,Pellois2022, Hossain2022, Zandbergen2022, Wang2022, Yang2022, Tan2022} 
or both sensors with magnetometers~\cite{Lee2009,Nazarahari2021,Kang2016,Abbasi-Kesbi2018,Duan20201,Peppoloni2013,Ruffaldi2014,Cockcroft2014,Fang2016,Zhang2011,Abbasi-Kesbi2017,Meng2016,Nazarahari2021b,Kun2011,McGrath2018,Mazomenos2014,Kawano2007,Fei2021,ConteAlcaraz2021,Sy2021,Sy2021b,Atrsaei2018,Slajpah2014,Wang2017,Butt2019,Schiefer2014,Saito2020,Yang2021,Fourati2017,Pathirana2018,Sim2013,Wouda2019,Nagaraj2021,Zhang2011b,Alvarez2016,Liu2010c,Li2022}.
Few studies join the accelerometer and magnetometer data~\cite{Zabat2015,Liu2010b}
and only one uses the gyroscope and magnetometer data~\cite{Butt2021}.

\begin{figure}[ht!]
    \centering
    \includegraphics[width=0.73\columnwidth]{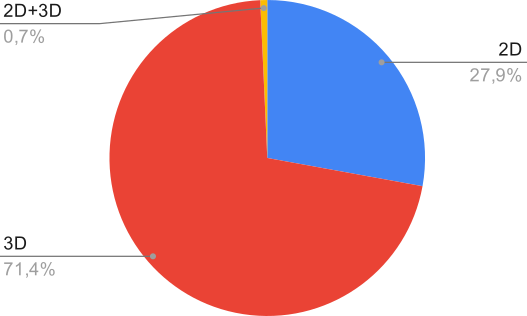}
    \caption{Type of estimations in term of their dimensionality, divided between 2D and 3D estimations.}
    \label{fig:2d3d}
\end{figure}

The 3D position of devices can be inferred from the IMU sensors information.
The combination of the three sensors, gyroscope, magnetometer and accelerometer includes information of the angular rate of motion and the references of the vector gravity and Earth magnetic field references. 
However, these 3D positions can also be estimated by sensor fusion techniques using different combinations of the three sensors in IMUs. 
Most studies give 3D estimations (\unit[$71.4$]{\%}), as shown in Fig.~\ref{fig:2d3d}.  
Conversely, only \unit[$1.4$]{\%} of works use  accelerometers and magnetometers and \unit[$25.2$]{\%} both sensors complemented with gyroscopes (see Fig.~\ref{fig:sensor}).
This fact is noticeable because the combination of the first two sensors is required to obtain the references to overcome the gyroscope drift and get accurate 3D estimations. 
It implies that the majority of algorithms that offer 3D space predictions propose methods for error reduction that do not rely on vector references.
In this way, the magnetic disturbances
that cause errors in the magnetic field measurements are avoided.

The 2D estimations include kinematic parameters in any plane perpendicular to the floor and the two angles with respect to the horizontal plane, in the frontal and sagittal planes. 
No doubt the 3D estimations are more complete since they provide information about the whole motion, even if it is mostly performed in one plane. 
That is the reason why  only \unit[$27.9$]{\%} of studies focus on obtaining estimations in the $2$D space.

Only one study adapts the estimation to 3D or 2D spaces, according to the motions~\cite{Choi2018}. 
In this proposal, the method gives 2D estimations based on the accelerometer data when the motion is mostly performed in one plane.
If deviations from this plane are detected, the method integrates the gyroscope data in order to provide 3D estimations.

\subsection{Application}
\label{sub_find:app}

\begin{figure*}[htbp!]
    \centering
    \includegraphics[width=0.7\textwidth]{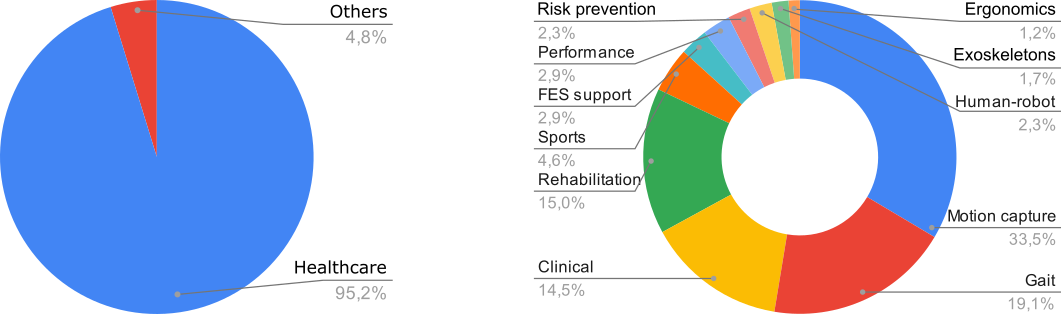}
    \caption{Analysis of the application of the studied works. Left: percentage of proposals whose aimed field is included in healthcare-related applications. Right: specific application or applications considered in proposals. Some works provide possible uses of their proposals, others focus on a specific application, such as gait, and others refer to the general motion capture field. FES refers to functional electrical stimulation and the research of human-robot interactions (HRI) is labeled as ``Human-robot''. 
    }
    \label{fig:applications}
\end{figure*}

Healthcare applications are the most common ones (\unit[$95.2$]{\%}) in the inertial motion analysis field (see Fig.~\ref{fig:applications}). These applications include motion capture or analysis, gait and clinical assessment, or rehabilitation. 
The aim of \unit[$33.5$]{\%} of studies is the motion capture in order to obtain information about human kinematics for the motion analysis or find possible diseases.  
Gait is the second most common application (\unit[$19.1$]{\%}) due to its relationship with cognitive impairments. 
The prevalence of use for the specific clinical assessment is similar,  
being the aim of the \unit[$14.5$]{\%} of works.
Rehabilitation and sports  are also worth mentioning because they are very motivating in research works (\unit[$19.6$]{\%} of studies).

\subsection{Monitored Motion Unit}
\label{sub_find:IMU}

\begin{figure}[ht!]
    \centering
    \includegraphics[width=0.73\columnwidth]{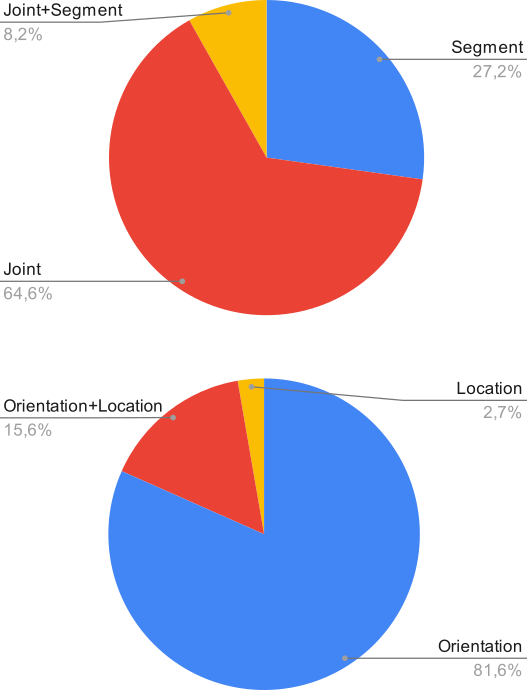}
    \caption{Monitored motion units and the obtained measurement. Top: percentage of studies that  measure each motion unit or their combination. Bottom: type of measurement, orientation and location.}
    \label{fig:or_loc_joint_segment}
\end{figure}

We analyze the anatomical unit measured in the reviewed works.
In this work, the anatomical units are called monitored motion units following the nomenclature of previous studies~\cite{lopez_nava_review}. 
We divide these monitored motion units into two groups: segments and joints.
Segments usually correspond to elements of the skeletal system, such as thighs (femur), and are modeled as a rigid-solid bodies. 
Joints are the unions between segments. 
The objective of \unit[$64.6$]{\%} of studies is to measure the motion of joints, whereas the \unit[$27.2$]{\%} of proposals focus on tracking segments and the remaining \unit[$8.2$]{\%} combine the monitoring of both monitored motions, segments and joints, as shown in Fig.~\ref{fig:or_loc_joint_segment}-top.

\begin{figure*}[htbp!]
    \centering
    \includegraphics[width=0.8\textwidth]{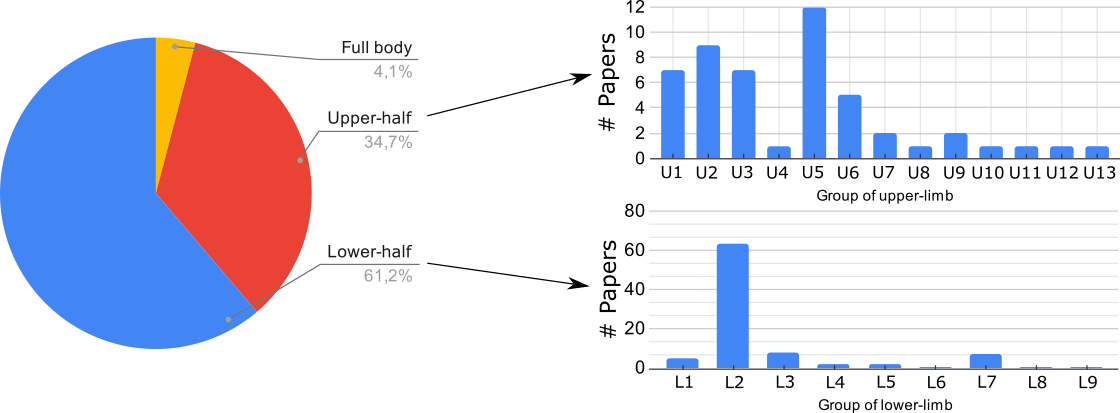}
    \caption{Anatomical monitored units. Left: location of the monitored units in the upper-or lower-half part of the body. Right: number of papers per combination of segments and/or joints monitored.}
    \label{fig:LU_limbs}
\end{figure*}

Studies focus more frequently on the lower-half (\unit[$61.2$]{\%}) of the body than on than upper-half (\unit[$34.7$]{\%}). 
Compared to the outcomes in the review of Lopez-Nava~\cite{lopez_nava_review}, this trend is in the most recent works different than in the previous ones. 
We found that recent research, dated on the last three years,  extend motion analysis to full-body monitoring, which is an important difference with the findings of previous sudies~\cite{lopez_nava_review}.
We consider full body if both upper-and lower-halves are monitored, which is made in the \unit[$4.1$]{\%} of studies.
Fig.~\ref{fig:LU_limbs}-left depicts the percentage of works that monitor each body half or the full body.

We study the groups of segments or joints included in each body half for a deeper analysis. 
We define the groups as sets of monitoring units.
We consider that studies focus on one of the groups if they estimate the orientation or location of one of the included monitored units. For example, one study that tracks the motion of wrists is included in the hand group, since we establish that the hand group includes the wrist among other motion units.

With respect to the upper-half of the body, we divide it into hand (hand, wrist and fingers), arm segments (arm and forearm), arm joints (shoulder, elbow and forearm twist), trunk (back, trunk and torso) and head and upper back (head/neck/scapula). 
The upper-half groups (\unit[$51$]{works}) are named as follows:
arm segments (U1), 
trunk (U2), 
arm joints and hand (U3),  
head and upper back, trunk, arm segments and arm joints (U4), 
arm joints (U5), 
head and upper back (U6), 
arm segments and joints (U7), 
trunk, arm joints and hand (U8), 
head and upper arm, trunk and arm (U9), 
head and upper arm and arm joints (U10), 
head and upper back and trunk (U11),
head and upper back and arm segments (U12) and
head and upper back alone (U13).
Fig.~\ref{fig:LU_limbs}-right shows the presence of the combination of these groups in the studied works.

The arm joints group, U5, is the one on which most works are focused (\unit[$12/51$]{studies}), followed by the trunk (\unit[$9/51$]{studies}), U2.
The next three frequent groups are the arm segments (\unit[$7/51$]{studies}), U1, 
the combination of arm and hand joints (\unit[$7/51$]{studies}), U3, and 
the head and upper back (\unit[$6/51$]{studies}), U6. 
The rest of the groups are only monitored in $1/51$ or \unit[$2/51$]{studies}, according to the case.
In this way, research works commonly focus on the study of the arms more than the other upper-half body structures.

With regard to the lower-half, we divide it into pelvis, leg segments (thigh and shin), leg joints (hip/knee/ankle) and feet. The names of the groups of their combinations are the following:
leg segments and feet (L1),
leg joints (L2), 
leg segments (L3), 
leg joints and feet (L4),
feet (L5),
pelvis, leg segments and leg joints (L6),
leg segments and joints (L7), 
pelvis and leg joints (L8)
and pelvis (L9).
Fig.~\ref{fig:LU_limbs}-right shows the number of works focused on each of these groups (the total number of works is $90$).

In the lower-half body, it is noticeable that most of the works focus on the leg joints, L2, is the object of monitoring of most works (\unit[$63/90$]{studies}).
These joints are commonly studied in multiple applications, being the most important the gait analysis because of its relevance 
in health assessment. 
It is worth mentioning the contrast of the number of studies about the L2 group in comparison with the L4 group, that combines the leg joints with the feet, meaning that the motion of feet joints is commonly discarded in the studies focused on the lower-limb joints.
Beside the leg joints, the following most studied groups that are also focused on legs are:
leg segments, L3  (\unit[$8/90$]{studies}), 
leg segments and joints, L7 (\unit[$7/90$]{studies}) and
leg segments and feet, L1 (\unit[$5/90$]{studies}).
The rest of the groups, that include 
leg joints and feet (L4),
feet (L5),
pelvis, leg segments and leg joints (L6)
and pelvis (L9) are studied in few works 
(\unit[$2/90$]{studies}).

The monitored units in these groups are measured regarding their  
orientation or location.  
Orientation refers to the rotation angles, which are commonly presented as Euler angles or quaternions. Locations refer to the spatial coordinates, so they are a measurement of distance.
Most of studies estimate the orientation of monitored units (\unit[$81.6$]{\%}), the \unit[$15.6$]{\%} give a combination of orientation and location of units and only the \unit[$2.7$]{\%} are focused only on providing locations. Fig.~\ref{fig:or_loc_joint_segment} shows the distribution of the measured magnitudes.

\subsection{Adopted Algorithms}
\label{sub_find:sensorfusion}

The algorithms used in the estimation of the kinematic parameters can be separated into five different groups:
integration,
vector observation,
sensor fusion filters,
ML techniques
and other methods.

\subsubsection{Sensor fusion filters}

Sensor fusion filters (SF in Table~\ref{tab:alg+sens+est}), including Kalman Filters (KF), particle filters (PF), and Complementary Filters (CF) are the algorithms most frequently used.
Specifically, KFs are still the algorithms that are employed the most in the inertial human motion analysis field, following the trend reported in previous studies~\cite{lopez_nava_review}.

The problem formulation of Bayesian filters  consists in the identification of the desirable estimations using a series of measurements observed over time containing statistical noise and different inaccuracies~\cite{simon2006optimal}. 
The inputs and observations form the knowledge on the system's behavior and both convey errors and uncertainties, namely: the measurement noise and the system errors.
These filters fuse the information of sensors with the knowledge of the system in two stages: the estimation stage and the update stage. 
The initial stage uses the information of the previous time instant to estimate the current state of the state vector.
The second stage updates these estimations using the measurements from the sensors.

The motion analysis includes proposals with Extended Kalman filter (EKF), KF and \emph{Unscented} Kalman filter (UKF), in descending order of frequency of use. 
KF is a sensor fusion technique that estimates the states of a linear system through the minimization of the variance of the estimation error~\cite{simon2006optimal}. 
KFs use a series of measurements observed over time and their statistical noise to produce estimates of unknown variables.
EKFs appeared because KFs are limited to linear systems, being their generalization to non-linear systems. 
EKFs assume that the non-linearities in the dynamic and the observation model are smooth, so they expand the state and observation functions in Taylor series and approximate in this way the next estimate of the state vector.
However, this approximation can introduce large errors in the true posterior mean and covariance of the variables, which may lead to the divergence of the filter. 
One of the possible solutions is the use of UKFs, 
whose distribution of their state vector is a set of sample points called \emph{sigma points}. 
Sigma points capture the actual mean and covariance of the Gaussian random variables and are obtained though the \emph{Unscented Transformation} (UT). 
The UT is a method for calculating the statistics of a random variable that suffers a nonlinear transformation. 
UKFs are an extension of UTs to the recursive estimation where the UT is applied to the augmented state vector.

EKFs are the KF variation that most  frequently  appear in motion analysis. 
EKFs are used for the sensor fusion of gyroscopes and accelerometers in order to estimate the joints orientation~\cite{Allen2017,Joukov2020,Young2010,Jakob2013,Caroselli2013,Joukov2015,Joukov2018}, joints orientation and location~\cite{Kitano2019,Joukov2019,Joukov2014,Lin2012,Salehi2020} and segment orientation~\cite{Music2008,Lin2013}.
Researchers also use EKFs to fuse gyroscope and accelerometer data with magnetometer measurements to estimate the orientation of joints~\cite{Slajpah2014, Saito2020} or segments~\cite{Nazarahari2021,Kang2016,Butt2019,Pathirana2018,Sy2021}.
EKFs are also combined with the Gauss-Newton algorithm to fuse the information of gyroscopes and accelerometers and estimate the orientation of joints~\cite{Weygers2020}.

Classical KFs are normally used  for the sensor fusion of gyroscope and accelerometer data to estimate the segment orientation~\cite{Teruyama2013,Ligorio2015,Liu2020,Mazza2012,Watanabe2014,Watanabe2015,Alvarado2017,Wang2017}, the joint orientation~\cite{Ligorio2020,Cooper2009,Zhou2010,Alizadegan2017, Mayorca-Torres2020},  the segment location~\cite{Xu2018}, and all of them, the segment and joint orientation and location~\cite{watanabe2011}. KFs have also been used in the fusion of gyroscope, accelerometer and magnetometer data to estimate segments orientation and location~\cite{Sy2021b, Li2022}.

UKFs appear  less frequently in the literature. 
UKFs are commonly used for the sensor fusion of gyroscopes, accelerometers and magnetometers to estimate the joints location~\cite{Atrsaei2018},
or orientation~\cite{McGrath2018,Zhang2011b},
the orientation of joints and segments~\cite{Mazomenos2014},
and the orientation and location of both elements, joints and segments~\cite{Peppoloni2013}.
Some works do not use the magnetometer information and only fuse the gyroscope and accelerometer data to estimate joints orientation~\cite{ElGohary2011,ElGohary2012}.

PFs are another modification of KFs for their use in non-linear systems~\cite{simon2006optimal}. 
PFs are close in functioning to UKFs but with a set of differences that approximates PFs to a generalization of UKFs.
PFs update the estimations with a randomly generated noise according to the \emph{prior} knowledge of the process noise Probability Density Function (PDF) instead of the update of the UKF that is deterministic.
Another difference with UKFs is that the number of particles in PFs is not related to the length of the state vector.
Finally, PFs estimates the PDF of the state instead of the mean and covariance, and it converges to the actual PDF as the number of particles increases. 

PFs are less popular than any other kind of KFs.
PFs are applied to fuse the measurements from gyroscopes, accelerometers and magnetometers to estimate the orientation of segments and joints~\cite{Zhang2011}.
PFs are combined with other KFs, such as EKFs, for the fusion of gyroscope and accelerometers measurements to estimate the orientation of segments~\cite{Villeneuve2017}.

CFs combine the information from different sensors by minimizing the mean-square-error instead the error covariance, which is minimized in KFs~\cite{Brown1972}.
CFs are used to fuse the measurements of gyroscopes, accelerometers and magnetometers to estimate the orientation of joints~\cite{Cockcroft2014, Abbasi-Kesbi2017, Fei2021}, their orientation and location~\cite{Abbasi-Kesbi2018, Fourati2017}, and their orientation together with the segments orientation~\cite{Yang2021}.
They are also used to estimate the joint either by combining the information of the gyroscope and the accelerometer orientation~\cite{Feldhege2015, Karunarathne2014, Chen2020,Meng2019, Sharma2017,Figueiredo2020}, or just only with the gyroscope~\cite{Allseits2018}.

Another alternative to KFs is the Weighted Fourier Linear Combiner (WFLC) filter, which is a model-based adaptive filter. WFLCs exploit the \emph{prior} knowledge of the signal shape and evolution over time, in those occasions when the motion performed is given~\cite{Bonnet2013a}. These filters are specially effective in periodic signals but adapt to variations between repetitions. 
Their applications to the human motion analysis include the use of the turn rate measurements to estimate the segment orientation~\cite{Grimpampi2013} and the combination of these data from the gyroscope with the accelerometer measurements to estimate the orientation of joints~\cite{Bonnet2013a}.

\subsubsection{Data science algorithms}
ML techniques represent the second group of algorithms that is applied most frequently for the estimation of the human kinematics. 
Furthermore, the supervised learning algorithms are the most widespread in recent years. 
Supervised learning is one of the most employed learning paradigm which
tries to discover the unknown function $f(\mathbf{x},\omega)$
that relates the input space  $X\subset \mathbb{R}^n$ (which, in this work, are the inertial measurements), with the output space $Y\subset\mathbb{R}$ (which describes the motion kinematics).
Each pair $(\mathbf{x}_i,{y}_i)$ is composed of the value of a set of $n$ predictive variables $\mathbf{x}_i=(x_1,\cdots,x_n)_i\in \mathbb{R}^n$ of the input space, which are measured by the IMUs, and its corresponding output value $y_{i}\in \mathbb{R}$, which are the target value of joint or segment orientation and location.
During the process called \emph{training}, supervised algorithms retrieve the map $f\in F$ from the provided training dataset $\mathcal{D}$, typically establishing an optimization problem that minimizes a \emph{loss function} $\mathcal{L}$.
Different parametric function spaces $F$ with different learning methods correspond to the existent variety of supervised methods, as described in depth in Table~\ref{tab:ML_detail}.


Gaussian Processes (GPs) are kernel-based probabilistic ML models. The Gaussian process is a kind of continuous random process $f(t)$, such that every finite set of random variables has a multivariate Gaussian distribution~\cite{rasmussen2003gaussian}. GP method estimates the output $y$ by introducing a set of latent variables $\lbrace f(t_k) \rbrace_{k=1}^n$ from a Gaussian process, and explicit link functions, $g(\cdot)$. 
Gaussian Process Latent Variable (GPLV) models are used with the gyroscope and accelerometer data to estimate the segment positions~\cite{Eom2014}.

Other classical ML methods are Decision Trees (DTs) and Support Vector Machines (SVMs). 
A DT is a classical ML method that builds a tree, a particular graph without cycles, by branching decision paths for each considered input variable to make the final classification~\cite{rokach2005decision}.
During the training process, databases are used to compute thresholds (the parameters in DTs) that better branch the input variable for optimizing a criterion, usually the best gain of information possible in the current node (optimizing the entropy), for a better prediction of the output variable.

SVMs are one of the most used ML method for classification~\cite{scholkopf2002learning,scholkopf2000new}.
It establishes an optimization problem to find the so-called \textit{support vectors}, those training data which are close to the separation hyperplane and maximize the \textit{soft margin}. Frequently, this method uses the kernel trick which consists in choosing an appropriate non-linear mapping $\boldsymbol\phi$ that maps input samples into a higher dimensional space where they are likely to be linearly separable. In regression problems, the support vectors are used to provide a continuous value trough a link function instead of classes.

However, these classical ML methods are less promising than artificial  Neural Networks (ANNs) in the human motion analysis field, as proved in~\cite{Figueiredo2020} for the correction of the joint angles initially obtained from sensor fusion filters. ANNs consist in a set of connected base units known as artificial neurons which emulate the biological neurons of animal brains~\cite{jain1996artificial}.
ANNs are usually organized in layers which interconnect themselves to create a huge variety of networks that try to represent the functional relation between the input and output variables.
ANNs have revolutionized the ML field due to their ability to model very complex non-linear input-output relations and their capacity to learn them from a huge amount of data.
The single Multi-Layer Perceptrons (MLP) were the first ANNs. 

In the inertial motion capture field, ANNs use the accelerometer data as inputs to estimate the segments orientation and location~\cite{Lim2020}, combine the gyroscope and accelerometer data to estimate joint orientations~\cite{Mundt2020,Lee2021,Bennett2013,Mundt2020b} or fuse the information of the three sensors integrated in IMUs to estimate the segment angles~\cite{Wouda2019}. 
Other specific types of ANNs merge the estimation of joint angles with gyroscopes and accelerometers, such as the general regression NNs~\cite{Findlow2008, Goulermas2008} or the Elman  Neural Networks~\cite{Tan2022}.

Deep  Neural Networks (DNNs) arise later than ANNs, and encompass a huge amount of modern network architectures with a high number of interconnected layers~\cite{alghifari2018speech}. The current technology allows massive computation during the training process, hence, new variety of interconnections and predictions in real time.
DNNs starts with the Convolutional Neural Networks (CNNs), a large sequence of convolutional layers configured in cascade where each layer computes the convolution operation  (see~\cite{oppenheim1997signals}) from the previous one. They are able to extract intrinsic local features, the called \textit{deep features}, which surpass the results of the classical ML methods.
VGG~\cite{simonyan2014very} and Residual Networks (RESNET)~\cite{RESNET} are famous CNNs included in this category. 
Most of DNNs including CNNs are feed-forward networks which means that the information flows forward and they do not include cycles.
However, DNNs also include recurrent networks which memorize internal states, frequently exploited for temporal sequence, such as the improved Recurrent Neural Network (RNN)~\cite{pearlmutter1989learning}, which evolved to the novel Long-Short Term-Memory (LSTM)~\cite{hochreiter1997long}, Gate Recurrent Unit (GRU)~\cite{cho2014learning} and nonlinear autoregressive neural network with exogenous inputs (NARX)~\cite{lin1996learning}.

Among the deep learning algorithms, 
LSTMs are the most utilized. 
LSTMs are made by a sequence of cells capable to keep previous states, specifically keep two kind of temporal information, the long and short-term memory. They have replaced the RNNs which suffer from the vanishing gradient problem during the training and include forget gates to quickly adapt to the new changes of data. 
These DNNs can  use just the information of accelerometers and the orientation of a set of body segments to estimate the whole-body posture~\cite{Zheng2021} 
or fuse the information of specific force with the turn rate to estimate the joint angles~\cite{Mundt20201, Hernandez2021,SharifiRenani2021,Rapp2021}. 
A less common approach includes the fusion of gyroscopes and magnetometers to estimate the  joint angles~\cite{Butt2021}.
LSTMs can also be used to estimate the orientation of the whole-body joints using the orientation obtained with sparse commercial sensors~\cite{Nagaraj2021}.
In~\cite{Rapp2021}, LSTMs are combined with CNN to estimate the joint angles. CNNs are also used to obtain the joint angles only using the accelerometer data~\cite{Gholami2020} or fusing the gyroscope and accelerometer data~\cite{Dorschky2020, Hossain2022, Yang2022}.
In~\cite{Mundt2021}, Mundt et al. made a comparison of these previous methods, CNNs and LSTMs, together with MLPs for the estimation of joint orientation. 
Using the information of gyroscopes and accelerometers, CNNs provided the most favorable metrics.  
Other RNNs are also used to estimate the joints orientation. 
To estimate the joint angles from gyroscopes and accelerometers, \cite{Tham2021} proposes a NARX; and \cite{ConteAlcaraz2021} also includes the magnetometer data with NARX and LSTMs.

\begin{figure}[htbp!]
    \centering
    \includegraphics[width=0.8\columnwidth]{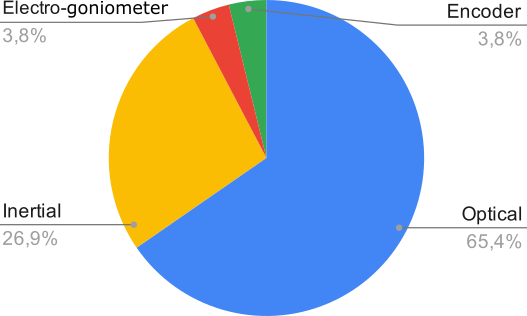}
    \caption{External sensors used to obtain the reference measurements for the training and test stages of the ML algorithms.}
    \label{fig:external_sensors_ML_training}
\end{figure}

The ML-based algorithms that are used for human motion analysis are supervised methods. These methods need training data, which must include  reference data of the parameter to estimate, i.e. the joint or segment orientation or location. In this review, we found 26 works that use reference data, that 
can be obtained from a stereophotogrammetric system (17/26)~\cite{Findlow2008,Goulermas2008,Eom2014, Mundt20201,Gholami2020,Dorschky2020, Lim2020, Mundt2020,Mundt2020b,Rapp2021,Hernandez2021,SharifiRenani2021,Mundt2021, Tham2021, Hossain2022, Yang2022, Tan2022},
electro-goniometer and encoders (2/26)~\cite{Bennett2013, Lee2021}
or inertial sensors (7/26)
~\cite{Wouda2019, Figueiredo2020, Zheng2021,  Nagaraj2021, ConteAlcaraz2021, Liang2021,Butt2021}.  Fig.~\ref{fig:external_sensors_ML_training} shows the percentage of use of the different external sensors for obtaining reference data.

The use of optical systems also allows the data generation in order to have data for the training and test of the algorithms and to increase the available data to train the models. These tasks can be performed with simulation software, e.g. OpenSim~\cite{opensim}, as in~\cite{Nagaraj2021, Hernandez2021, SharifiRenani2021}, 
by applying kinematic relationships from the stereophotogrammetric measurements, as in~\cite{Bagala2012, Mundt20201, Dorschky2020, Rapp2021,SharifiRenani2021, Mundt2020, Mundt2020b}, 
or with data augmentation techniques~\cite{Mundt2020}

\begin{table*}[htbp]
\centering
\caption{Outcomes of the analysis of inputs used in the ML algorithms, algorithms applied in each work, and outputs aimed as targets with the capture system employed.
\textbf{SF} : specific force, \textbf{TR}: turn rate, \textbf{OR}: orientation, \textbf{Cap.}: capture, \textbf{SNN}: Shallow  Neural Network, \textbf{DNN}: Deep  Neural Network, vs.: \emph{versus}
}
\begin{tabular}{l|c|c|c|l|l|l|c} 
\hline
\multirow{2}{*}{\textbf{Work}} & \multicolumn{3}{c|}{\textbf{Inputs}}                                                                       & \multirow{2}{*}{\textbf{Specifications of inputs}}                                                                                                                                                                              & \multirow{2}{*}{\textbf{Algorithm}}                                                                                                   & \multicolumn{2}{c}{\textbf{Outputs}}                                               \\ 
\cline{2-4}\cline{7-8}  & \textbf{SF}     & \textbf{TR} & \textbf{OR}  &   &       & \textbf{Magnitudes}                                           & \textbf{Cap. system}      \\ 
\hline
\cite{Findlow2008}           & \checkmark       & \checkmark  &     & Time instant  & GRNN                     & Joint angle                                          & Optical             \\ 
\hdashline
\cite{Goulermas2008}         & \checkmark  & \checkmark  &     & Time instant            & \begin{tabular}[c]{@{}l@{}}GRNN\\Aux. Sim. Info.\end{tabular}                                               & Joint angle                                          & Optical             \\ 
\hdashline
\cite{Bennett2013}            & \checkmark                   & \checkmark  &     & Time instant            & ANN      & Joint angle                                          & Other \\ 
\hdashline
\cite{Eom2014}               & \checkmark      & \checkmark  &     & Sparse IMUs                     & GPLVM      & Whole-body posture                                    & Optical             \\ 
\hdashline
\cite{Wouda2019}             & \checkmark    &    & \checkmark   & \begin{tabular}[c]{@{}l@{}}Sparse IMUs\\\unit[$5$]{samples} (\unit[$0.55$]{s})  windows\end{tabular}     &  SNN   \& DNN 
        & 
\begin{tabular}[c]{@{}l@{}}Segment location\\Whole-body posture\end{tabular} 
& Inertial            \\ 
\hdashline
\cite{Figueiredo2020}        &    & \checkmark  & \checkmark   & Joint angle from Bayesian filters       & \begin{tabular}[c]{@{}l@{}l@{}} NN (vs.)\\ SVM (vs.)\\ DT
\end{tabular}            & Joint angle        & Inertial            \\ 
\hdashline
\cite{Mundt20201}            & \checkmark     & \checkmark  &     & \begin{tabular}[c]{@{}l@{}}Simulated inertial data\\Different IMU combination +~PCA\\From $100$ to \unit[$1000$]{time-steps} \end{tabular}      & LSTM   & Joint angle        & Optical             \\ 
\hdashline
\cite{Gholami2020 }          & \checkmark    &    &     & \begin{tabular}[c]{@{}l@{}} \unit[$600$]{ms} windows \\Sparse IMUs for leg kinematics\end{tabular}           & 1D-CNN              & Joint angle      & Optical             \\ 
\hdashline
\cite{Dorschky2020}          & \checkmark    & \checkmark  &     & \begin{tabular}[c]{@{}l@{}}Experimental and simulated data combined\\ \unit[$100$]{time-steps} for each walking cycle\end{tabular}    & CNN   & \begin{tabular}[c]{@{}l@{}}Joint angle \\ Joint moment \\ GRF  \end{tabular}   & Optical             \\ 
\hdashline
\cite{Lim2020}              & \checkmark  & \checkmark  &     & Time instant          & ANN  & \begin{tabular}[c]{@{}l@{}}Joint angle \\ Joint moment \\ GRF  \end{tabular}                            & Optical             \\ 
\hdashline
\cite{Mundt2020}             & \checkmark  & \checkmark  &     & \begin{tabular}[c]{@{}l@{}l@{}}Simulation of IMUs \\
Data augmentation\\\unit[$10$]{frames} windows\\Full gait cycles\end{tabular}   & ANN     & \begin{tabular}[c]{@{}l@{}}Joint angle \\ Joint moment  \end{tabular}                                  & Optical             \\ 
\hdashline
\cite{Mundt2020b}            & \checkmark     & \checkmark  &     & \begin{tabular}[c]{@{}l@{}}Simulated inertial data\\Segmentation into gait cycles\end{tabular}    & ANN   & \begin{tabular}[c]{@{}l@{}}Joint angle \\ Joint moment \end{tabular}                               & Optical             \\ 
\hdashline
\cite{Zheng2021}             & \checkmark   &    &  \checkmark/\xmark & 
\begin{tabular}[c]{@{}l@{}}Sparse IMUs\\Window of \unit[$50$]{samples} $=$ \unit[$0.50$]{s}\end{tabular}           & LSTM      & Whole-body posture                             & Inertial            \\ 
\hdashline
\cite{Rapp2021}              & \checkmark                                                                                      & \checkmark  &     & \begin{tabular}[c]{@{}l@{}l@{}l@{}}Simulated inertial data\\Calibration for alignment\\Marker trajectories filtered\\\unit[$1$]{s} \& \unit[$2$]{s} windows\end{tabular}                                                                                          & \begin{tabular}[c]{@{}l@{}l@{}}1D-CNN (vs.)\\LSTM\\ \& Optimizer\end{tabular} & \begin{tabular}[c]{@{}l@{}} Joint angle \\ time series \end{tabular}                              & Optical             \\ 
\hdashline
\cite{Hernandez2021}         & \checkmark & \checkmark  &     &  \unit[$100$]{samples}  (\unit[$100$]{Hz})  windows   & \begin{tabular}[c]{@{}l@{}}CNN\\LSTM\end{tabular}                                                                                                                    & \begin{tabular}[c]{@{}l@{}} Joint angle \\ time series \end{tabular}                                 & Optical             \\ 
\hdashline
\cite{SharifiRenani2021}     & \checkmark                                                                                      & \checkmark  &     & \begin{tabular}[c]{@{}l@{}}Experimentally measured IMU data,\\Simulated inertial data\\ \unit[$200$]{time-steps} \\Time wrapping of strides as inputs\end{tabular} & LSTM                                                                                                                         & \begin{tabular}[c]{@{}l@{}} Joint angle \\ time series \end{tabular}                                & Optical             \\ 
\hdashline
\cite{Butt2021}              & \checkmark                                                                                      & \checkmark  &     & \begin{tabular}[c]{@{}l@{}}Sparse IMUs\\Synthetic and experimental data\\ \unit[$300$]{frames}  windows\end{tabular}                                                                                                                                & LSTM                                                                                                                         & \begin{tabular}[c]{@{}l@{}} Joint angle \\ time series \\ Whole-body posture \end{tabular}                               & Inertial            \\ 
\hdashline
\cite{Nagaraj2021}           &                                                                                        &    & \checkmark   & \begin{tabular}[c]{@{}l@{}}IK outputs as biRNN inputs\\+Joint parmeter\\Sparse IMUs\\  \unit[$300$]{time-steps}  windows\end{tabular}                                                                                                   & LSTM                                                                                                                         & Whole-body posture                            & Inertial            \\ 
\hdashline
\cite{Mundt20201}             & \checkmark                                                                                      & \checkmark  &     & \begin{tabular}[c]{@{}l@{}}Experimental and simulated data\\ \unit[$101$]{frames}  windows\end{tabular}                                                                                                                               & \begin{tabular}[c]{@{}l@{}}MLP\\LSTM\\CNN\end{tabular}                                                                       & \begin{tabular}[c]{@{}l@{}} Joint angle \\ time series \end{tabular}                                & Optical             \\ 
\hdashline
\cite{Lee2021 }              & \checkmark                                                                                      & \checkmark  & \checkmark   & Time instant                                                                                                                                                                                                              & ANN                                                                                                                          & \begin{tabular}[c]{@{}l@{}}Joint angle \\ Walking speed  \end{tabular}                           & Other             \\ 
\hdashline
\cite{Tham2021}              & \checkmark & \checkmark  &     & Time instant                                                                                                                                                                                                              & \begin{tabular}[c]{@{}l@{}}NARX (vs.) \\CF   \end{tabular}                                                                                                       & Segment angle                                  & Optical             \\ 
\hdashline
\cite{ConteAlcaraz2021}      &  \checkmark & \checkmark  & \checkmark   & \begin{tabular}[c]{@{}l@{}}Sparse IMUs for leg kinematics \\Gait cycle segmentation\end{tabular}                                                                                                                                              & \begin{tabular}[c]{@{}l@{}}NARX \\LSTM   \end{tabular}                                                                                                                                                 & Joint angle                                          & Inertial            \\ 
\hdashline
\cite{Liang2021}            & \checkmark          & \checkmark  &     & \begin{tabular}[c]{@{}l@{}}Sparse IMUs for knee angle\\Short intervals \unit[$0.05$]{s}\end{tabular}                                                                                                             & LSTM                                                                                                                         & Joint angle                                          & Inertial       \\
\hdashline
\cite{Hossain2022}            & \checkmark          & \checkmark  &     & \begin{tabular}[c]{@{}l@{}}Sparse IMUs for leg kinematics\\Evaluation of classifier combinations
 \end{tabular}                                                                                                             & \begin{tabular}[c]{@{}l@{}} CNN \\ RNN \end{tabular}                                                                                                                         & \begin{tabular}[c]{@{}l@{}} Joint angle \\ time series \end{tabular}                                          & Optical       \\
\hdashline
\cite{Yang2022}            & \checkmark          & \checkmark  &  \checkmark  & \begin{tabular}[c]{@{}l@{}}Initial gyro integration for joint orientation estimation
 \end{tabular}                                         & Elman NN      & Joint position                                   & Optical       \\
\hdashline
\cite{Tan2022}            & \checkmark          & \checkmark  &     & Gait variations                      & LSTM      &  \begin{tabular}[c]{@{}l@{}} Joint angle \\ time series \end{tabular}                                         & Optical       \\
\hline
\end{tabular}\label{tab:ML_detail}
\end{table*}

\subsubsection{Other algorithms}
\label{subsubsec:otheralgorithms_methods}

Over the years of research on motion analysis with inertial sensors, proposals have been based on various algorithms other than sensory fusion filters and data science methods.
These proposals cover from the integration of the gyroscope data to estimate the joint orientation~\cite{Niijima2014,An2012, Mohammadzadeh2015, Watanabe2016,Choi2018}, to its combination with the direct use of the data from accelerometers to estimate the orientation of joints~\cite{Tadano2013, Kumar2018, Seel2014} and segments~\cite{Charry2011, Williamson2001, Baten1996,Falbriard2020}, and to estimate the orientation and location of segments~\cite{Pellois2022, Zandbergen2022}.
The measurements from gyroscopes and accelerometers are also used directly to obtain the orientation and location of joints~\cite{Ohtaki2001} and segments~\cite{Liu2009}, and to estimate the orientation of both joints and segments~\cite{Hu2014,Flores-Morales2016, Wang2022}.
The information of the three sensors in the IMU are also directly used for the estimation of the segments orientation~\cite{Schiefer2014}. 






Different works exploit the observation of the gravity vector by the accelerometer for the estimation of the orientation of joints~\cite{Willemsen1991,Lee2010,Das2014,Bakhshi2011,Sun2016}, segments~\cite{Zehr2021} or both~\cite{Laidig2017}.
Other works also use the data  of the magnetometer to estimate thee joint orientation and location~\cite{Watson20211}, or combine this information with the measurements of accelerometers to obtain the joint orientation~\cite{Zabat2015}.
The gravity vector can also be observed by eliminating of the linear acceleration of the motions, which can be estimated from the turn rate measurements~\cite{Takeda2009}.





Besides the gyroscope integration and the direct observation of vectors, 
the measurements from IMU sensors can be combined through the use of virtual sensors. 
The use of virtual sensors consists in the estimation of the measurements that a sensor would obtain if it was located in the joint.
This measurement projection is commonly performed because it is not possible to place the sensors in the joints.
This is commonly used to simulate the measurements in joints whereas the IMUs are placed in segments.
This approach is used to combine the gyroscope and accelerometer measurements to estimate only the joint orientation~\cite{Dejnabadi2005, Liu2009,Fasel2018}, or both the joint orientation and location~\cite{Liu2010}. 
Another application of virtual sensors is to combine the turn rate, specific force and magnetic field to obtain these magnitudes in joints to estimate their orientation~\cite{Kun2011,Alvarez2016,Liu2010c, Kawano2007} 
and also not considering the measurements from gyroscope for the joint orientation estimation~\cite{Liu2010b}.

Another common method used is the gradient descent.
Gradient descent is applied to obtain the joint orientation by using the measurements from gyroscopes~\cite{Muller2017} and combined with the specific force measurements~\cite{Yin2021,Ding2020,Weygers2020}. 
This approach also allows the gathering of the measurements of the three sensors to estimate the segment orientation~\cite{Lee2009}.

The remaining proposals use a wide variety of methods and approaches to monitor the measurement units.
These methods include probabilistic graphical models~\cite{Ruffaldi2014}, which are employed with the gyroscope, accelerometer and magnetometer measurements to estimate the orientation of joints,  smoothing algorithms~\cite{Molnar2018}, bidirectional low-pass filters~\cite{Bagala2012}, least squares~\cite{Cehajic2015, Bonnet2016}, optimization techniques~\cite{Dorschky2019,Djuric-Jovicic2012,Zhou2008} and modified iterative algorithms~\cite{Ma2015}.
The latest worth mentioning approach consists in the double-sensor difference based algorithm~\cite{Liu2009c}, that combines the measurements from two accelerometers placed on the same segment with the knowledge of their positions with respect to the joint.

\subsubsection{Approaches for error reduction}

This section describes the main approaches found in the literature in order to reduce the errors in the estimation of kinematic parameters. First, we focus on the explanation of the proposals based on biomechanical constraints and then we summarize the approaches for error reduction based on the properties of the inertial sensors and their motions.

\begin{figure}[htpb]
    \centering
    \includegraphics[width=.8\columnwidth]{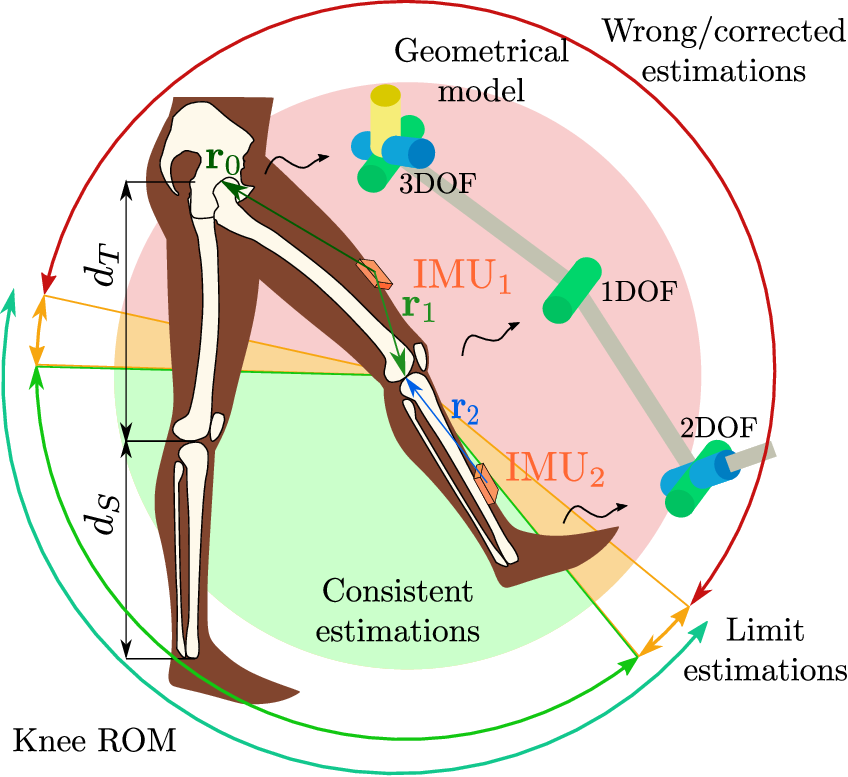}
    \caption{Scheme of the biomechanical constraints commonly implemented in the kinematic models for the inertial motion analysis. It includes the geometrical model with a reduced number of DOF in the knee and ankle joints and the limitations with respect to the ROM, using the knee as example. The lengths and IMU-joint vector used in the soft constraints are labeled as $d_S$, $d_T$ and $\boldsymbol{r}_i$ with $i=0,1,2$, respectively. }
    \label{fig:leg_constraint}
\end{figure}

Biomechanical constraints are a promising resource to improve the inertial human motion analysis.
A common approach is to model the rotations of different joints with different Degrees Of Freedom (DOF), which are depicted with cylinders in the geometrical model in Fig.~\ref{fig:leg_constraint}.
The three rotational DOF recorded by IMUs can be modeled as one or two DOF joints, according to the possible anatomical motions. 

In the literature, the knee is assumed to be a hinge joint with just one DOF (flex-extension, see Fig.~\ref{fig:leg_constraint})~\cite{Yin2021,Feldhege2015,McGrath2018,Chen2020,Sy2021,Sy2021b,Lee2021,Seel2014,Joukov2014,Joukov2015,Joukov2018,Sharma2017,Bonnet2016,Liu2010c,Joukov2019} because its internal rotation allows a negligible Range Of Motion (ROM).
Some works also consider the knee internal rotation together with its flex-extension, so the modeled joint presents two DOF~\cite{Cooper2009}.
The same approach can be used to simplify the ankle orientation estimation, using only the knee-flex extension rotation, so one DOF~\cite{Sy2021b} or also including the internal rotation, resulting in two DOF~\cite{Joukov2019}, as Fig.~\ref{fig:leg_constraint} depicts.
Less commonly used due to its actual DOF and complexity, hips are modeled as joints with two DOF~\cite{Sharma2017,Liu2010c}.

This DOF reduction can be also applied to upper-limbs.
Elbows can be assumed to have only two DOF, gathering the elbow flex-extension rotation and the forearm internal-external rotation~\cite{Alizadegan2017,ElGohary2012,Zhang2011b,Alvarez2016,ElGohary2011, Muller2017,Peppoloni2013,Ruffaldi2014, Yang2021, Joukov2019} or considering only the elbow flex-extension~\cite{Zhang2011}.
Wrist can be also modeled as one DOF joint~\cite{Peppoloni2013,Ruffaldi2014} or allowing a second rotation with two DOF~\cite{Alvarez2016}.

Another approach related with the simplification of motions to a lower amount of DOF is to model the motions as if they occur in  one or two planes.  
This approach reduces the $3$D space in, at least, one dimension.
Different motions, such as gait or squats, can be approximated as $2$D in the sagittal plane~\cite{ConteAlcaraz2021,Kumar2018, Figueiredo2020} or with the combination of the sagittal and coronal planes~\cite{Alvarado2017,Watanabe2014}.

The separation of motions in the DOF available for the joints allows another restriction based on the joint anatomical ROM.
This constraint is based on the correction of the estimations that are not consistent with the anatomically possible ROM per DOF of joints. 
As depicted in Fig.~\ref{fig:leg_constraint}, the ROM of a joint, in this case the knee, includes the consistent estimations and the estimations on the limit of the ROM. The values of angles outside this range are wrong estimations of the algorithm and the objective is to detect and correct them.
This approach can be found in several proposals in the literature~\cite{Duan20201,Sy2021b,Lin2012,Bonnet2016, Yang2021,Charry2011}.

Another common approach for the error reduction in 
most of biomechanical models is to take into account the anatomic parameters, such as the joints location with respect to the sensors or the segments length~\cite{Ohtaki2001,Abbasi-Kesbi2018,Dejnabadi2005,Allen2017,Kun2011,Liu2009,Kawano2007,Bagala2012,Laidig2017,Liu2009,Weygers2020,Watanabe2016,Takeda2009,Kitano2019,Atrsaei2018,Music2008,Slajpah2014,Lin2012,Lin2013,Ma2015,Fasel2018,Willemsen1991,Bonnet2016,Djuric-Jovicic2012,Liu2009c,Joukov2014,Liu2010,Caroselli2013,Bonnet2013a,Villeneuve2017,Zhou2010,Yang2021,Joukov2015,Joukov2018,Alizadegan2017,ElGohary2012,Kumar2018,Alvarez2016,ElGohary2011,Zhou2008,Salehi2020,Liu2010b,Liu2010c,Mundt2020b}.
Fig.~\ref{fig:leg_constraint} shows these parameters, labeled as $\boldsymbol{r}_i$, where $i=0,1,2$, $d_S$ and $d_T$, respectively.
The IMU-joint vector and the segment length are used to impose that the relationship between magnitudes has to be consistent with the anatomy of participants and the location of sensors on the body.
These constraints can be applied by the relationship between the velocity in the common joint between two segments and the turn rate measured by the gyroscopes of the corresponding IMUs using Eq.~\eqref{eq:vomega}, see
Fig.~\ref{fig:leg_constraint}.
\begin{equation}
    \boldsymbol{v}_{\textit{knee}}=\boldsymbol\omega_{\textit{IMU}_1}\times \boldsymbol{r}_1 = \boldsymbol\omega_{\textit{IMU}_2}\times \boldsymbol{r}_2\label{eq:vomega}
\end{equation}

An alternative approach is to relate the linear acceleration suffered by the IMUs with the linear acceleration in the common joint between segments. This approach requires the consideration of the gravity influence from the specific force measured by accelerometers. If the gravity influence is eliminated, Eq.~\eqref{eq:aomegaacc} can be applied with the derivation of the turn rate.
\begin{multline}
    \boldsymbol a_{\textit{knee}} = \boldsymbol{a}_{\textit{IMU}_1} +
    \dot{\boldsymbol{\omega}}_{\textit{IMU}_1} \times \boldsymbol{r}_1 +
    \boldsymbol\omega_{\textit{IMU}_1} \times \left( \boldsymbol\omega_{\textit{IMU}_1} \times \boldsymbol{r}_1
    \right)=\\
    a_{\textit{IMU}_2} +
    \dot{\boldsymbol{\omega}}_{\textit{IMU}_2} \times \boldsymbol{r}_2 +
    \boldsymbol\omega_{\textit{IMU}_2} \times \left( \boldsymbol\omega_{\textit{IMU}_2} \times \boldsymbol{r}_2
    \right)
    \label{eq:aomegaacc}
\end{multline}

The IMU-joint vectors combined with the segment lengths or the joint-joint vectors are commonly used to estimate the kinematic of chains of segments. 
This is frequently performed with the Denavit-Hartenberg (D-H) notation, that uses four angular and distance parameters to relate reference frames with the links of spatial kinematic chains~\cite{dhn}. 
In order to apply the D-H convention, one reference frame is defined for each DOF included in the biomechanical model.
The axes of the consecutive reference frames, $i-1$ and $i$, must follow two rules: the x$_i$ axis must be perpendicular to z$_{i-1}$ and the x$_i$ axis must intersect with z$_{i-1}$. In this way, the transformation matrix $T_{i-1,i}$ detailed in Eq.~\eqref{RT} defines the transformation between consecutive frames.
\begin{equation}\label{RT}
T_{i-1,i} = 
\begin{bmatrix}
\cos\theta_i & \text{-}\cos\beta_i\hspace{2pt}\sin\theta_i & \sin\beta_i\hspace{2pt}\sin\theta_i & r_i\cos\theta_i \\
\sin\theta_i & \cos\beta_i\hspace{2pt}\cos\theta_i & \text{-}\sin\beta_i\hspace{2pt}\cos\theta_i & r_i\sin\theta_i\\
0 & \sin\beta_i & \cos\beta_i & d_i\\
0&0&0&1
\end{bmatrix}
\end{equation}
Where $\theta_i$ is the angle between $x_{i-1}$ and $x_i$ axis, about the $z_{i-1}$ axis and  $\beta_i$ is the angle between $z_{i-1}$ and $z_{i}$ axis, about the $x_{i}$ axis.
This transformation of consecutive frames allows the estimation of the forward kinematics of a chain of joints by using Eq.~\eqref{eq:vomega} and Eq.~\eqref{eq:aomegaacc}, as performed in~\cite{Bonnet2016,Peppoloni2013,Ruffaldi2014,Xu2018,Lin2012}. 

The location of joints and the segment lengths are not imposed as the limitation of DOF or ROM, which directly models or corrects the estimations. 
However, they are used to impose constraints in the measured magnitudes. 
For that reason, the restrictions forced through these conditions are commonly known as \emph{soft constraints}. 
It is worth mentioning that the errors in the estimation of the IMU-joint vector directly influence the estimation of joint angles that use these soft constraints~\cite{Bonnet2016}. 

Other corrections found in the literature that are not related with biomechanical properties include the following:
modeling the bias of sensors or including them in the state vector~\cite{Teruyama2013,Lee2008181,Kang2016,Molnar2018,Abbasi-Kesbi2018,Cockcroft2014,Zhang2011, Ligorio2020,Muller2017,Watanabe2014,Bagala2012, Watanabe2015, Liu2009,Young2010,Cehajic2015, Wang2017, Cooper2009, Zhou2010, Lee2010, Nazarahari2021, Williamson2001, Bakhshi2011, Zabat2015, Chen2020, Falbriard2020, Fasel2018}; 
calibrating this bias~\cite{Lee2010, Liu2009, Ligorio2020, Fang2016, Peppoloni2013, Alvarado2017, Tadano2013, Kumar2018, Alvarez2016, Fei2021};
low-pass filtering the recorded signals~\cite{Dejnabadi2005, Allseits2018, Choi2018,Feldhege2015, Falbriard2020,Grimpampi2013, Pathirana2018, Yang2021, Lee2021}; or optimizing them~\cite{Dorschky2019},
updating the estimations when a direct observation of gravity is available or with its dynamic compensation~\cite{Ohtaki2001,Liu2020, Hu2014, Flores-Morales2016, Charry2011, Butt2019, Baten1996, Schiefer2014, Sy20201}  
or the zero-angle during the zero-turn rate time instants~\cite{Allseits2018, Karunarathne2014, Hu2014};
modeling the disturbances in the magnetometer and gyroscope~\cite{Kang2016,Watson20211}; using virtual sensors, 
discriminating the quasi-static and dynamic motions~\cite{Choi2018};
eliminating the errors from the soft tissue artifacts~\cite{Ligorio2020,Meng2016} and, 
in the case of Kalman filters, the optimization of the  Kalman parameters~\cite{Teruyama2013,Mazza2012,Lin2012}.

\subsection{Participants of the study}
\label{sub_find:participants_app}

\begin{figure}[ht!]
    \centering
    \includegraphics[width=0.73\columnwidth]{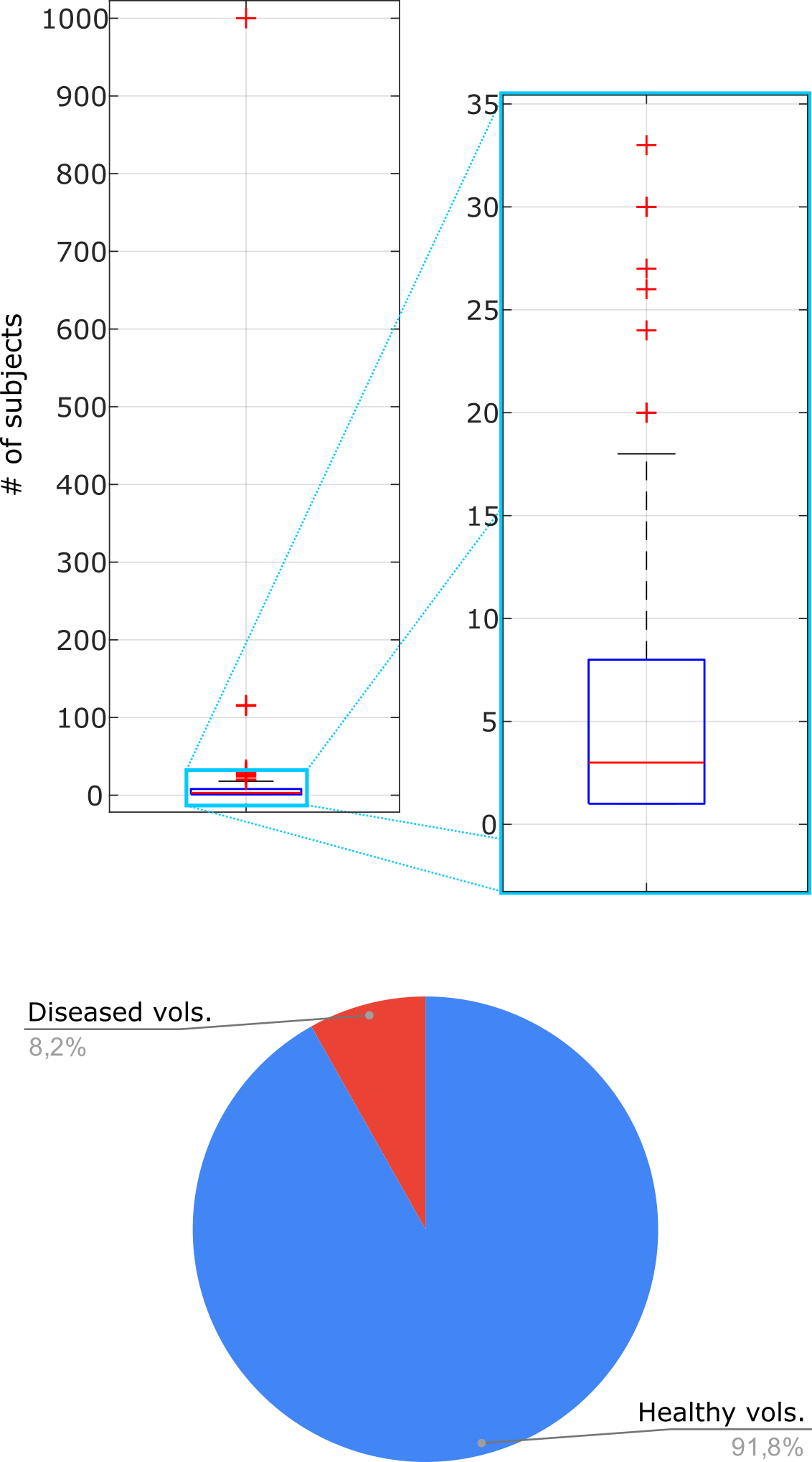}
    \caption{Characteristics of the population of study in the human motion analysis literature. Top: percentage of works that evaluates their proposal in each number of participants. Bottom: percentage of works that considers population with (Diseased vols.) or without (Healty vols.) disease.}
    \label{fig:subjects}
\end{figure}

This work also analyzes the number of subjects that participate in the studies to validate the methods.
Fig.~\ref{fig:subjects}-top shows the \emph{boxplot} of the distribution of subjects in the studies analyzed in this work ($147$). 
The boxplot presents the $1^{\text{st}}$, $2^{\text{nd}}$, $3^{\text{rd}}$ and $4^{\text{th}}$ quartile of the studied subjects together with the \emph{outliers}.
In this work, the outliers represent the punctual studies that test their proposals in more than \unit[$18$]{subjects} (\unit[$8/147$]{studies}).
According to Fig.~\ref{fig:subjects}-top, most  results provided in the studies correspond to a population of less than \unit[$10$]{subjects}. 
It is worth mentioning than the median is \unit[$3$]{subjects} per study, what makes the results hardly generalizable to all population. 
Furthermore, more than one third of studies test their proposals with only one person (\unit[$34.7$]{\%}). 

The studies that validate their proposal with the highest amount of volunteers commonly test ML-based algorithms. 
This amount of volunteers is required by these algorithms because they work with a high amount of data in order to develop generalizable models. 
However, a \unit[$65.4$]{\%} of these studies use the data from the optical systems to generate simulated inertial data, as shown in Fig.~\ref{fig:external_sensors_ML_training}.

Studies analyze volunteers with or without diseases related with the motor system. We  
assume that if there is no statement about whether unhealthy people is included, the studied population is healthy or with no illness that affects the performance of motions.
In this way, only a small percentage (\unit[$8.2$]{\%}) of proposals are tested on population with these motor limitations (see Fig.~\ref{fig:subjects}-bottom). This is remarkable since most proposals claim healthcare applications among their possible uses as seen in section~\ref{sub_find:app} (\unit[$95.2$]{\%}).

\subsection{Validation Systems and Evaluation Metrics}
\label{sub_trends:validation}

\begin{figure}[ht!]
    \centering
    \includegraphics[width=0.8\columnwidth]{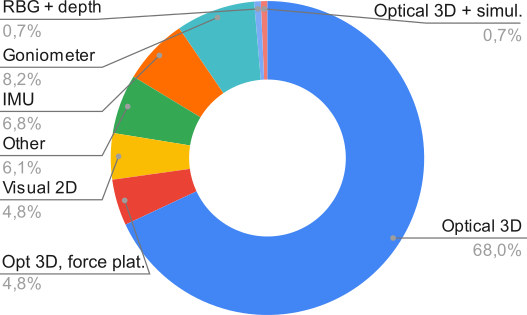}
    \caption{Validation system used to assess the proposals. Force platforms and simulations are abbreviated as force plat. and simul., respectively.}
    \label{fig:val_sys}
\end{figure}

For the validation of the reviewed works, researchers use different systems, as shown in Fig.~\ref{fig:val_sys}.
The gold standard is the $3$D optical motion capture system, such as the commercial Vicon~\cite{vicon} or Optitrack~\cite{optitrack}, being the most widely employed.
This system is commonly used for the validation of proposals (\unit[$68.0$]{\%}) and, sometimes (\unit[$4.8$]{\%}), in combination of force platforms or with a simulation software (\unit[$0.7$]{\%}).
$2$D optical systems that can obtain a reference in the image plane are also used (\unit[$4.8$]{\%} of the works). In some works, the $2$D optical systems are combined with depth sensors (\unit[$0.7$]{\%}). 

Another approach to validate the  algorithms that is worth mentioning is the use of output values of commercial inertial systems that provide highly accurate measurements, as done in the \unit[$6.8$]{\%} of studies.  
IMUs of the Xsens commercial brand are the most frequently used for the validation of proposals~\cite{xsens}. Four of the eleven works that validate their algorithms against inertial sensors outputs use these sensors~\cite{Pellois2022,Nagaraj2021,Figueiredo2020,Wouda2019}, whereas the remaining seven works use IMUs of seven different brands.

A less common solution includes the use of analog and electronic goniometers (\unit[$8.2$]{\%}). Other validation systems, which in combination sum a \unit[$6.8$]{\%} of works, include different programs of motion simulation, encoders and potentiometers.

The accuracy metric reported most frequently for the validation of proposals is the root mean square error (RMSE). In some works, the correlation coefficient or the mean absolute error (MAE) are also provided. For the case of angles measurement, the studied works report a RMSE between \unit[$2.59$]{\textdegree} and \unit[$7.67$]{\textdegree}. Even if in average most of studies provide similar metrics, it is worth mentioning that the RMSE range of ML methods is between \unit[$2.48$]{\textdegree} and \unit[$5.70$]{\textdegree}, whereas the RMSE provided by the classical methods is between \unit[$2.24$]{\textdegree} and \unit[$7.80$]{\textdegree}.
These results prove that ML methods are promising approaches in the human motion analysis field in spite of their limitations related to the data availability.

\section{Discussion}
\label{sec:discussion}

After the previous in-depth study, we discuss the review findings in terms of the general trends and the future guidelines in the inertial motion monitoring field. 

\subsection{General Trends}

This paper analyzes the current state and the research trends in the inertial motion analysis field.
The analyzed works show the interest in developing an alternative to the \textit{gold standard} system, based on cameras, due to their high cost and the required space of use. 
As a consequence, the research focused on developing an IMU-based system for the human motion analysis has increased over the last years, as seen in Fig.~\ref{fig:years}.

The sensors integrated in IMUs, accelerometer, gyroscope and magnetometer, are fused in different ways in the analyzed works, as shown in Fig.~\ref{fig:sensor}. The fusion of gyroscopes and accelerometers is more common than the use of  magnetometers, being one of the main differences with respect to the findings in previous reviews~\cite{lopez_nava_review}. 
However, the use of magnetometers is spreading during the last year with \unit[$10$]{works}.
Bayesian filters and trigonometric approaches are the ones that most frequently employ the data from the magnetometer. 
ML proposals barely rely on the measurements of the magnetic field and only use them in the training step of the algorithms.
In this way, the works focused on ML proposals are not limited by magnetic disturbances. 

The common objective in \unit[$72.4$]{\%} of studies (see Fig.~\ref{fig:2d3d}) is to obtain $3$D kinematic parameters.
The $2$D estimations are useful in human motion analysis because some movements can be simplified as motions in one plane, e.g. knee or elbow flex extension or even gait and squats. 
However, these estimations can miss relevant information in motions, about correctness or symptoms of motion-related diseases. 
Obtaining the complete kinematic information is specially important in healthcare applications, that is considered in the \unit[$95.2$]{\%} of works (see Fig.~\ref{fig:applications}). 

Most of the analyzed works (\unit[$33.5$]{\%}, as shown in Fig.~\ref{fig:applications}) mention the generic motion capture field for human motion analysis, closely followed by the gait evaluation, as aimed applications for their work. 
That reflects the interest of developing more affordable and user-friendly alternatives to the optical systems, as previously discussed. 
Consequently, the analyzed works propose algorithms to monitor frequently the orientation of joints, which is commonly measured by stereophotogrammetric systems, such as Vicon~\cite{vicon}.
The \unit[$64.6$]{\%} of works focus on joints and the \unit[$81.6$]{\%} on the estimation of the orientation. These percentages imply a great advance in the direction of inertial solutions for the human motion analysis, especially compared to the trends reported in~\cite{lopez_nava_review}, where most works studied the orientation of segments. 

Another interesting analysis is focused on the distribution of works with respect to the analyzed body part, divided into the upper-and lower-halves.
The \unit[$61.2$]{\%} of works (see Fig.~\ref{fig:LU_limbs}) study the lower-half of the body, and most of them focus on the leg joints, which is also consistent with the trend of gait evaluation besides the motion analysis.
Conversely, only the \unit[$34.7$]{\%} (see Fig.~\ref{fig:LU_limbs}) studies the upper-half, that include arms and trunk, which are difficult to monitor due to the DOF and the complexity of joints as shoulders or neck.
The results in Fig.~\ref{fig:LU_limbs} mean that during the last years, the research has been focused mostly on the lower-half of the body, the opposite of what happened in~\cite{lopez_nava_review}, that most of the reviewed works analyzed upper-limbs. 
However, monitoring this upper-half of the body is crucial for the evaluation of motions, being specially important in the rehabilitation of cognitive alterations or illness, such as strokes.
The remaining \unit[$4.1$]{\%}  (see Fig.~\ref{fig:LU_limbs}) of proposals are aimed at monitoring the whole-body posture, which are the most complete approach for the human monitoring. 
Even though the gait analysis is commonly performed by monitoring the lower-limbs, upper-limbs are also important to study relevant features, as balance, in clinical assessments.
The rising interest in considering the full body is also another noteworthy difference compared to previous findings. 

For the full-body monitoring, ML methods are specially attractive. 
Sensor fusion algorithms use one IMU per segment to monitor the whole-body posture or model biomechanical relationships between segments to reduce this number with different constraints.
Conversely, the approach in ML-based proposals focused on the whole-body posture is the optimization of the number of devices with the use of the so-called \emph{sparse} IMUs, as in~\cite{Eom2014,Wouda2019,Zheng2021,Butt2021,Nagaraj2021}.
This approach is also used to monitor specific limbs, as legs, reducing the number of sensors~\cite{Gholami2020,ConteAlcaraz2021,Liang2021}.

The biomechanical approaches for error reduction can restrict motions and might not be generalized for populations with motor related diseases. For instance, the ROM of joints can be different in people with anomalous physical abilities. 
Likewise, the assumption of a number of DOF can miss relevant information about motions out of the main directions.
Also, the knowledge of the segment length or the location of the sensors on the body is not always available in practical applications.
Different  IMU-joint calibration methods have been proposed to address this limitation.
The first approach is to obtain an average location of joints with respect to the sensors, which has been validated for the upper-and lower-limbs~\cite{Crabolu2017,sgv_memea2019}, respectively, but require specific calibration motions.
The second method consists in estimating an adaptive position vector, considering the changes of the location of IMUs due to soft tissue artifacts~\cite{Frick2018,Frick2018a,sgv_arved,sgv_memea2020}, which has been validated for the calibration of hips performing leg circles~\cite{sgv_arved}.
These proposals assume that the joints are fixed, but it is not the case in all the activities in the daily living.
In~\cite{lee2022_adaptcalib}, the calibration of moving joints with soft tissue artifacts is addressed.

As in previous findings~\cite{lopez_nava_review}, the sensor fusion algorithms are employed more commonly than other approaches. 
However, their use during  the last decade remains stable (around \unit[$24$]{papers}), whereas the use of ML techniques has increased from \unit[$4$]{papers} to \unit[$18$]{papers}. 
These ML techniques provide a slight improvement in the accuracy metrics, referred to a reduction of the maximum RMSE in \unit[$2$]{\textdegree} in angle measurements. However, none study analyzed in this work includes a research that makes a fair comparison using common data to test both approaches.

Sensor fusion filters and data science algorithms differ in terms of their computational costs. Computational time varies among the different methods depending on their implementations. Sensor fusion filters are faster than data science methods which require more calculations, specially, DNN-based ones whose number of parameters is superior. Also, ML-and DNN-based methods usually demand more memory than the sensor fusion solutions, specially in their training stage, making their implementations more expensive.

ML algorithms are more robust to variation in the intrinsic noise of the sensors with which they are trained. 
In addition, their robustness can increase by generating synthetic data to which more noise models are added. 
Conversely, the sensor fusion algorithms include parameter tuning to adapt them to the sensors used, e.g. the covariance matrix of KFs.
Thus, sensor fusion algorithms would require a previous study to estimate these sensor-dependent matrices.

ML methods require a high amount of reference data to be trained. Two alternative trends are followed in order to generate reference data: \emph{1)} to simulate the inertial data from the optical data to use them as inputs or \emph{2)} to use the orientation data obtained by commercial systems as reference. 
In the first case, the simulation of inertial data might not present the intrinsic  errors of IMUs, whereas by using inertial data as reference, it presents an error around \unit[$0.5$]{\textdegree} depending on the commercial brand, which is less accurate than optical systems. 

With regard to the validation, new reference systems  have appeared during the last years. Among them, we find $2$D visual systems, encoders and computational models.
The $3$D optical systems are still the ones most frequently used (\unit[$68.0$]{\%} of studies, see Fig.~\ref{fig:val_sys}).
However, the use of this validation system entails the limitation of testing the proposals in daily activities and alternative validation methods should be investigated~\cite{weygers2020_review}.

The reviewed studies generally analyze a low amount of participants for the validation of the algorithms.
This limitation was detected in~\cite{lopez_nava_review} and still remains in recent works. 
Most studies test their results only in one volunteer, and the average of study subjects is \unit[$4$]{participants}. It makes the proposals hardly to generalize for the whole population. 
In those studies that more participants are involved, the inertial data are simulated from the optical data or their reference consists in the orientation outputs obtained from the IMUs, including the errors previously indicated.

Most studies analyze healthy participants. That is noticeable since most studies consider healthcare applications as possible uses of their proposals. However, only a few of them (\unit[$8.2$]{\%}) test their proposals on subjects with motor related diseases.

\subsection{Future Advancements and Developments}

This review highlights a set of clear trends.
The studies describe the motions in the $3$D space more frequently than reducing them to planar motions. This is crucial to describe complex motions that can be performed during the daily life, so it is required for an out-of-the-lab analysis.
Furthermore, the reduction of the gait or simpler motions such as knee flex-extension to a plane eliminates relevant information about these motions.

The reviewed works focus on the lower-limbs, specifically on the orientation of the hip, knee and ankle. Future research should include upper-limbs or even focus on the development of whole-body posture monitoring for a complete description of motions. 
In this line of work, the proposal of sparse-IMUs utilization, is promising to decrease the number of sensors in use, which is required for the motion analysis in all environments.
Moreover, the monitoring of complex joints, such as shoulders or hips, which are usually modeled as $3$-DOF joints, should include all their DOF for a proper kinematic analysis.

With regard to the algorithms in use, the current trend moves from the Bayesian filters, which we consider the classical ones, to ML algorithms, specially into deep learning algorithms. 
For the development of these novel proposals, more data with an accurate reference are required, as described in~\cite{sgv2022_database,IMU_optic_gait_BBDD}, in order to avoid the use of data from IMUs and simulations from the optical systems as ground truth. 
One of the main limitations of the biomechanical constraints found in the literature is their generalization of use in wide and varied populations, where the constraints based on ROM and DOF exclude people with motor diseases. 
In this way, new proposals should be adaptable to the  populations under study.
Also, alternatives to obtain the IMU-joint vector based on inertial devices
are needed in order to make suitable the proposals that exploit the biomechanical relationships in out-of-the-lab environments.

Common data with inertial measurements and its reference are needed in order to obtain a fair comparison of the existent and new proposals. 
In that line of research, future proposals are required to be validated on a larger number of volunteers than is currently the case. 
This also should ensure the variability of motions and not be focused on the gait.

\section{Conclusions}
\label{sec:conclusions}

This work has reviewed the studies focused on the human motion analysis based on IMUs.
The date of publication of the reviewed papers is not limited, so we provide an overview of the proposals since the first study to the current date.
This overview summarizes the algorithms, the combination of sensors, the anatomical units monitored,  
the subjects of study and the validation approaches in the research of inertial monitoring. 
The review also focuses on the studies of the last decade, so we analyze the last trends in this research field. 
Most of the analyzed works focus on obtaining the $3$D estimation of the kinematics of lower-limb joints, presenting a lack of studies of the upper-half of the body.
The  Bayesian filters are still the most used methods, but their trend is to be applied less frequently, whereas the ML algorithms are being used now with a higher incidence. 
This review includes the description of 
the main algorithms used with their inputs and outputs for a better understanding of the existent methods.
In this way, we  show that, nowadays, these groups of algorithms present also differences in the selected sensors: Bayesian filters tend to use more the magnetometer and try to compensate its limitations, but ML algorithms commonly rely only on gyroscopes and accelerometers.
Both groups of algorithms present also differences in the range of accuracy,  obtaining slightly lower maximum errors by using ML methods.
This work  also analyzes the proposed approaches for error reduction, highlighting the need of proposals suitable for all the population and of IMU-joint calibration methods.
Finally, this work remarks the requirement of testing future proposals on a highly number of subjects, that help to create common databases that allow the comparison among the existent and new proposals.
 
\appendices

\appendices

\section{Tables of the data extracted}\label{sec:appendix}

\onecolumn
\begin{landscape}
\begin{longtable}{p{9cm} |p{.7cm} p{.6cm} p{.8cm} p{.45cm}| p{.5cm} p{.5cm} p{.5cm}  p{.5cm} | p{.8cm} P{0.5cm}  P{0.45cm}  P{0.45cm}  P{0.45cm} }
\caption{Relevant details related with the implemented algorithms, the sensors in use and the estimations of the selected studies. With respect to the FA acronyms, SF: sensor fusion, ML: machine learning, OA: other algorithms. Other acronyms: BC: biomechanical constraints, ANT: anatomical parameters in use, OC: other constraints, ML learns: refers to that theses ML-based algorithms learn for real motions so they are constraint to anatomical joint limits and EST: refers to the use of anatomical or soft constraints after estimate the required parameters, GS: gyroscope sensor, AS: accelerometer sensor, MS: magnetometer sensor, OS: other sensors for training, EST: type of estimation, ANG: angle, DIS: displacement or position, JNT: joint and SGM: segment.  }
\label{tab:alg+sens+est} \\
\hline
\multirow{2}{*}{\textbf{Title}}   & \multicolumn{4}{c|}{\textbf{Algorithms}}
& \multicolumn{4}{c|}{\textbf{Sensors}}    & \multicolumn{5}{c}{\textbf{Estimation characteristics}}                                     
\\
   & \textbf{FA}         & \textbf{BC} & \textbf{ANT}  & \textbf{OC}   & \textbf{GS} & \textbf{AS} & \textbf{MS} & \textbf{OS}                     & \textbf{EST}   & \textbf{ANG} & \textbf{DIS} & \textbf{JNT} & \textbf{SGM}               \\ 
\endfirsthead
\hline
\multirow{2}{*}{\textbf{Title}}   & \multicolumn{4}{c|}{\textbf{Algorithms}}                                                                               & \multicolumn{4}{c|}{\textbf{Sensors}}    & \multicolumn{5}{c}{\textbf{Estimation characteristics}}                                     
\\
   & \textbf{FA}         & \textbf{BC} & \textbf{ANT}  & \textbf{OC}   & \textbf{GS} & \textbf{AS} & \textbf{MS} & \textbf{OS}                     & \textbf{EST}   & \textbf{ANG} & \textbf{DIS} & \textbf{JNT} & \textbf{SGM}               \\ 
\endhead
\hline
A basic study on variable-gain Kalman filter based on angle error calculated from acceleration signals for lower limb angle measurement with inertial sensors~\cite{Teruyama2013}                                               & SF           &     &           & \checkmark   & \checkmark  & \checkmark  &    &                        & 2D    & \checkmark   &     &     & \checkmark                 \\ 
\hdashline
A Comparison of Three Neural Network Approaches for Estimating Joint Angles and Moments from Inertial Measurement Units~\cite{Mundt2021}                                                                                     & ML & \checkmark   & ML learns & \checkmark   & \checkmark  & \checkmark  &    & \checkmark                      & 3D    & \checkmark   &     & \checkmark   &                   \\ 
\hdashline
A fast quaternion-based orientation optimizer via virtual rotation for human motion tracking~\cite{Lee2009}                                                                                                                & OA           &     &           & \checkmark   & \checkmark  & \checkmark  & \checkmark  &                        & 3D    & \checkmark   &     &     & \checkmark                 \\ 
\hdashline
A Full-State Robust Extended Kalman Filter for Orientation Tracking during Long-Duration Dynamic Tasks Using Magnetic and Inertial Measurement Units~\cite{Nazarahari2021}                                                       & SF          &     &           & \checkmark   & \checkmark  & \checkmark  & \checkmark  &                        & 3D    & \checkmark   &     &     & \checkmark                 \\ 
\hdashline
A Human Motion Tracking Algorithm Using Adaptive EKF Based on Markov Chain~\cite{Kang2016}                                                                                                                                  & SF        &     &           & \checkmark   & \checkmark  & \checkmark  & \checkmark  &                        & 3D    & \checkmark   &     &     & \checkmark                 \\ 
\hdashline
A method for gait analysis in a daily living environment by body-mounted instruments~\cite{Ohtaki2001}                                                                                                                        & OA          & \checkmark   & \checkmark         &     & \checkmark  & \checkmark  &    &                        & 2D    & \checkmark   & \checkmark   & \checkmark   &                   \\ 
\hdashline
A Method for Lower Back Motion Assessment Using Wearable 6D Inertial Sensors~\cite{Molnar2018}                                                                                                                                & OA           &     &           & \checkmark   & \checkmark  & \checkmark  &    &                        & 3D    & \checkmark   &     &     & \checkmark                 \\ 
\hdashline
A Miniature Sensor System for Precise Hand Position Monitoring~\cite{Abbasi-Kesbi2018}                                                                                                                                             & SF           & \checkmark   & \checkmark         &     & \checkmark  & \checkmark  & \checkmark  &                        & 3D    & \checkmark   & \checkmark   & \checkmark   &                   \\ 
\hdashline
A new approach to accurate measurement of uniaxial joint angles based on a combination of accelerometers and gyroscopes~\cite{Dejnabadi2005}                                                                                     & OA           & \checkmark   & \checkmark         &     & \checkmark  & \checkmark  &    &                        & 2D    & \checkmark   &     & \checkmark   &                   \\ 
\hdashline
A new quaternion-based kalman filter for human body motion tracking using the second estimator of the optimal quaternion algorithm and the joint angle constraint method with inertial and magnetic sensors~\cite{Duan20201} & SF           & \checkmark   &           &     & \checkmark  & \checkmark  & \checkmark  &                        & 3D    & \checkmark   &     &     & \checkmark                 \\ 
\hdashline
A nonlinear dynamics-based estimator for functional electrical stimulation: Preliminary results from lower-leg extension experiments~\cite{Allen2017}                                                                        & SF          & \checkmark   & EST       &     & \checkmark  & \checkmark  &    &                        & 3D    & \checkmark   &     & \checkmark   &                   \\ 
\hdashline
A novel 7 degrees of freedom model for upper limb kinematic reconstruction based on wearable sensors~\cite{Peppoloni2013}                                                                                                        & SF          & \checkmark   & \checkmark         &     & \checkmark  & \checkmark  & \checkmark  &                        & 3D    & \checkmark   & \checkmark   & \checkmark   & \checkmark                 \\ 
\hdashline
A Novel Application of Flexible Inertial Sensors for Ambulatory Measurement of Gait Kinematics~\cite{Yin2021}                                                                                                              & OA           & \checkmark   & EST       &     & \checkmark  & \checkmark  &    &                        & 3D    & \checkmark   &     & \checkmark   &                   \\ 
\hdashline
A novel approach to motion tracking with wearable sensors based on Probabilistic Graphical Models~\cite{Ruffaldi2014}                                                                                                           & ML          & \checkmark   & \checkmark         &     & \checkmark  & \checkmark  & \checkmark  &                        & 3D    & \checkmark   &     & \checkmark   &                   \\ 
\hdashline
A novel complimentary filter for tracking hip angles during cycling using wireless inertial sensors and dynamic acceleration estimation~\cite{Cockcroft2014}                                                                     & SF        &     & EST       &     & \checkmark  & \checkmark  & \checkmark  &                        & 3D    & \checkmark   &     & \checkmark   &                   \\ 
\hdashline
A novel data glove for fingers motion capture using inertial and magnetic measurement units~\cite{Fang2016}                                                                                                                 & SF           &     &           & \checkmark   & \checkmark  & \checkmark  & \checkmark  &                        & 3D    & \checkmark   &     &     & \checkmark                 \\ 
\hdashline
A novel hierarchical information fusion method for three-dimensional upper limb motion estimation~\cite{Zhang2011}                                                                                                           & SF           & \checkmark   &           & \checkmark   & \checkmark  & \checkmark  & \checkmark  &                        & 3D    & \checkmark   &     & \checkmark   & \checkmark                 \\ 
\hdashline
A Novel Kalman Filter for Human Motion Tracking With an Inertial-Based Dynamic Inclinometer~\cite{Ligorio2015}                                                                                                                 & SF           &     &           &     & \checkmark  & \checkmark  &    &                        & 3D    & \checkmark   &     &     & \checkmark                 \\ 
\hdashline
A novel method for estimating knee angle using two leg-mounted gyroscopes for continuous monitoring with mobile health devices~\cite{Allseits2018}                                                                              & SF           &     &           & \checkmark   & \checkmark  &    &    &                        & 3D    & \checkmark   &     & \checkmark   &                   \\ 
\hdashline
A novel method of using accelerometry for upper limb FES control~\cite{Sun2016}                                                                                                                                            & OA        &     &           &     &    & \checkmark  &    &                        & 2D    & \checkmark   &     &     & \checkmark                 \\ 
\hdashline
A novel sensor-based assessment of lower limb spasticity in children with cerebral palsy~\cite{Choi2018}                                                                                                                    & OA        &     &           & \checkmark   & \checkmark  & \checkmark  &    &                        & 2D+3D & \checkmark   &     & \checkmark   &                   \\ 
\hdashline
A patient-centric sensory system for in-home rehabilitation~\cite{Abbasi-Kesbi2017}                                                                                                                                                 & SF           &     &           &     & \checkmark  & \checkmark  & \checkmark  &                        & 3D    & \checkmark   &     & \checkmark   &                   \\ 
\hdashline
A preliminary test of measurement of joint angles and stride length with wireless inertial sensors for wearable gait evaluation system~\cite{watanabe2011}                                                                      & SF           &     &           &     & \checkmark  & \checkmark  &    &                        & 2D    & \checkmark   & \checkmark   & \checkmark   & \checkmark                 \\ 
\hdashline
A study of gait analysis with a smartphone for measurement of hip joint angle~\cite{Niijima2014}                                                                                                                               & OA          &     &           &     & \checkmark  &    &    &                        & 2D    & \checkmark   &     & \checkmark   &                   \\ 
\hdashline
A Wearable Human Motion Tracking Device Using Micro Flow Sensor Incorporating a Micro Accelerometer~\cite{Liu2020}                                                                                                         & SF           &     &           & \checkmark   & \checkmark  & \checkmark  &    &                        & 2D    & \checkmark   &     &     & \checkmark                 \\ 
\hdashline
A Wearable Magnetometer-Free Motion Capture System: Innovative Solutions for Real-World Applications~\cite{Ligorio2020}                                                                                                        & SF           &     &           & \checkmark   & \checkmark  & \checkmark  &    &                        & 3D    & \checkmark   &     & \checkmark   &                   \\ 
\hdashline
Accuracy Improvement on the Measurement of Human-Joint Angles~\cite{Meng2016}                                                                                                                                               & OA          &     &           & \checkmark   & \checkmark  & \checkmark  & \checkmark  &                        & 3D    & \checkmark   &     & \checkmark   &                   \\ 
\hdashline
Accuracy of a custom physical activity and knee angle measurement sensor system for patients with neuromuscular disorders and gait abnormalities~\cite{Feldhege2015}                                                            & SF           & \checkmark   &           &     & \checkmark  & \checkmark  &    &                        & 3D    & \checkmark   &     & \checkmark   &                   \\ 
\hdashline
Adaptive Gain Regulation of Sensor Fusion Algorithms for Orientation Estimation with Magnetic and Inertial Measurement Units~\cite{Nazarahari2021b}                                                                                & SF           &     &           & \checkmark   & \checkmark  & \checkmark  & \checkmark  &                        & 3D    & \checkmark   &     &     & \checkmark                 \\ 
\hdashline
Alignment-Free, Self-Calibrating Elbow Angles Measurement Using Inertial Sensors~\cite{Muller2017}                                                                                                                            & OA          & \checkmark   & EST       &     & \checkmark  &    &    &                        & 3D    & \checkmark   &     & \checkmark   &                   \\ 
\hdashline
Ambulatory estimation of knee-joint kinematics in anatomical coordinate system using accelerometers and magnetometers~\cite{Kun2011}                                                                                       & OA           & \checkmark   & \checkmark         &     & \checkmark  & \checkmark  & \checkmark  &                        & 3D    & \checkmark   &     & \checkmark   &                   \\ 
\hdashline
Ambulatory measurement and analysis of the lower limb 3D posture using wearable sensor system~\cite{Liu2009}                                                                                                               & OA           & \checkmark   & \checkmark         &     & \checkmark  & \checkmark  &    &                        & 3D    & \checkmark   &     & \checkmark   &                   \\ 
\hdashline
An adaptive complementary filter for inertial sensor based data fusion to track upper body motion~\cite{Karunarathne2014}                                                                                                           & SF           &     &           &     & \checkmark  & \checkmark  &    &                        & 3D    & \checkmark   &     & \checkmark   &                   \\ 
\hdashline
An auto-calibrating knee flexion-extension axis estimator using principal component analysis with inertial sensors~\cite{McGrath2018}                                                                                          & SF          & \checkmark   & EST       &     & \checkmark  & \checkmark  & \checkmark  &                        & 3D    & \checkmark   &     & \checkmark   &                   \\ 
\hdashline
An inertial human upper limb motion tracking method for robot programming by demonstration~\cite{Pellois2022}                                                                                                  & OA          &    &  \checkmark         &   \checkmark  & \checkmark  & \checkmark  &    &                        & 3D    & \checkmark   &     & \checkmark   &                   \\                
\hdashline
An inertial sensor system for measurements of tibia angle with applications to knee valgus/varus detection~\cite{Hu2014}                                                                                                  & OA          &     &           &     & \checkmark  & \checkmark  &    &                        & 3D    & \checkmark   &     & \checkmark   & \checkmark                 \\ 
\hdashline
An instance-based algorithm with auxiliary similarity information for the estimation of gait kinematics from wearable sensors~\cite{Goulermas2008}                                                                               & ML          & \checkmark   &           & \checkmark   & \checkmark  & \checkmark  &    & \checkmark                      & 3D    & \checkmark   &     & \checkmark   &                   \\ 
\hdashline
An investigation into the accuracy of calculating upper body joint angles using MARG sensors~\cite{Mazomenos2014}                                                                                                                & SF          &     &           &     & \checkmark  & \checkmark  & \checkmark  &                        & 3D    & \checkmark   &     & \checkmark   & \checkmark                 \\ 
\hdashline
An optimized Kalman filter for the estimate of trunk orientation from inertial sensors data during treadmill walking~\cite{Mazza2012}                                                                                       & SF           &     &           & \checkmark   & \checkmark  & \checkmark  &    &                        & 2D    & \checkmark   &     &     & \checkmark                 \\ 
\hdashline
Analysis of a mobile system to register the kinematic parameters in ankle, knee, and hip based in inertial sensors~\cite{Flores-Morales2016}                                                                                          & OA          &     &           &     & \checkmark  & \checkmark  &    &                        & 3D    & \checkmark   &     & \checkmark   & \checkmark                 \\ 
\hdashline
Analyzing 3D knee kinematics using accelerometers, gyroscopes and magnetometers~\cite{Kawano2007}                                                                                                                             & OA       & \checkmark   & \checkmark         &     & \checkmark  & \checkmark  & \checkmark  &                        & 3D    & \checkmark   &     & \checkmark   &                   \\ 
\hdashline
Angle measurements during 2D and 3D movements of a rigid body model of lower limb: Comparison between integral-based and quaternion-based methods~\cite{Watanabe2014}                                                           & SF           & \checkmark   &           & \checkmark   & \checkmark  & \checkmark  &    &                        & 2D    & \checkmark   &     &     & \checkmark                 \\ 
\hdashline
Artificial neural networks in motion analysis—applications of unsupervised and heuristic feature selection techniques~\cite{Mundt20201}                                                                                       & ML         & \checkmark   & \checkmark         & \checkmark   & \checkmark  & \checkmark  &    & \checkmark                      & 3D    & \checkmark   &     & \checkmark   &                   \\ 
\hdashline
Biomechanical Ambulatory Assessment of 3D Knee Angle Using Novel Inertial Sensor-Based Technique~\cite{Tham2021}                                                                                                            & ML         & \checkmark   & ML learns & \checkmark   & \checkmark  & \checkmark  &    &                        & 3D    & \checkmark   &     & \checkmark   &                   \\ 
\hdashline
Calibrated 2D angular kinematics by single-axis accelerometers: From inverted pendulum to N-Link chain~\cite{Bagala2012}                                                                                                      & OA           & \checkmark   & \checkmark         &     &    & \checkmark  &    &                        & 2D    & \checkmark   &     &     & \checkmark                 \\ 
\hdashline
Closed-chain pose estimation from wearable sensors~\cite{Joukov2019}                                                                                                                                                          & SF          & \checkmark   & NO?       & \checkmark   & \checkmark  & \checkmark  &    &                        & 3D    & \checkmark   & \checkmark   & \checkmark   &                   \\ 
\hdashline
CNN-Based Estimation of Sagittal Plane Walking and Running Biomechanics From Measured and Simulated Inertial Sensor Data~\cite{Dorschky2020}                                                                                    & ML          & \checkmark   &           & \checkmark   & \checkmark  & \checkmark  &    & \checkmark                      & 2D    & \checkmark   &     & \checkmark   &                   \\ 
\hdashline
Comparison of angle measurements between integral-based and quaternion-based methods using inertial sensors for gait evaluation~\cite{Watanabe2015}                                                                             & SF           &     &           &     & \checkmark  & \checkmark  &    &                        & 2D    & \checkmark   &     &     & \checkmark                 \\ 
\hdashline
Data-Driven Reconstruction of Human Locomotion Using a Single Smartphone~\cite{Eom2014}                                                                                                                                    & ML         & \checkmark   &           & \checkmark   & \checkmark  & \checkmark  &    & \checkmark                      & 3D    &     & \checkmark   &     & \checkmark                 \\ 
\hdashline
DeepBBWAE-Net: A CNN-RNN Based Deep SuperLearner For Estimating Lower Extremity Sagittal Plane Joint Kinematics Using Shoe-Mounted IMU Sensors In Daily Living~\cite{Hossain2022} & ML          &             & ML learns    &   & \checkmark           & Y           &             & \checkmark & 2D           & \checkmark            &              & \checkmark            &               \\
\hdashline
Deriving kinematic quantities from accelerometer readings for assessment of functional upper limb motions~\cite{Laidig2017}                                                                                                   & OA          & \checkmark   & \checkmark         &     &    & \checkmark  &    &                        & 3D    & \checkmark   &     & \checkmark   & \checkmark                 \\ 
\hdashline
Design and validation of an ambulatory inertial system for 3-D measurements of low back movements~\cite{Charry2011}                                                                                                           & OA          & \checkmark   &           & \checkmark   & \checkmark  & \checkmark  &    &                        & 3D    & \checkmark   &     &     & \checkmark                 \\ 
\hdashline
Detecting absolute human knee angle and angular velocity using accelerometers and rate gyroscopes~\cite{Williamson2001}                                                                                                           & OA          &     &           & \checkmark   & \checkmark  & \checkmark  &    &                        & 2D    & \checkmark   &     &     & \checkmark                 \\ 
\hdashline
Development of a body joint angle measurement system using IMU sensors~\cite{Bakhshi2011}                                                                                                                                      & OA       &     &           & \checkmark   & \checkmark  & \checkmark  &    &                        & 3D    & \checkmark   &     & \checkmark   &                   \\ 
\hdashline
Development of a wearable glove system with multiple sensors for hand kinematics assessment~\cite{Fei2021}                                                                                                                 & SF           &     &           & \checkmark   & \checkmark  & \checkmark  & \checkmark  &                        & 3D    & \checkmark   &     & \checkmark   &                   \\ 
\hdashline
Development of a wearable sensor system for quantitative gait analysis~\cite{Liu2009b}                                                                                                                                      & OA          & \checkmark   & \checkmark         & \checkmark   & \checkmark  & \checkmark  &    &                        & 2D    & \checkmark   & \checkmark   &     & \checkmark                 \\ 
\hdashline
Digital inclinometer for joint angles measurements with a real-time 3D-animation~\cite{Zabat2015}                                                                                                                            & OA        &     &           & \checkmark   &    & \checkmark  & \checkmark  &                        & 3D    & \checkmark   &     & \checkmark   &                   \\ 
\hdashline
Drift-Free 3D Orientation and Displacement Estimation for Quasi-Cyclical Movements Using One Inertial Measurement Unit: Application to Running~\cite{Zandbergen2022}                             & OA          &             &              & \checkmark                & \checkmark          & \checkmark           &             &             & 3D           & \checkmark            & Y            &              & \checkmark             \\ 
\hdashline
Drift-Free and Self-Aligned IMU-Based Human Gait Tracking System with Augmented Precision and Robustness~\cite{Chen2020}                                                                                                    & SF           & \checkmark   &           & \checkmark   & \checkmark  & \checkmark  &    &                        & 3D    & \checkmark   &     & \checkmark   &                   \\ 
\hdashline
Drift-Free Foot Orientation Estimation in Running Using Wearable IMU~\cite{Falbriard2020}                                                                                                                                        & OA         &     & EST       & \checkmark   & \checkmark  & \checkmark  &    &                        & 3D    & \checkmark   &     &     & \checkmark                 \\ 
\hdashline
Drift-Free Inertial Sensor-Based Joint Kinematics for Long-Term Arbitrary Movements~\cite{Weygers2020}                                                                                                                         & SF       & \checkmark   & EST       & \checkmark   & \checkmark  & \checkmark  &    &                        & 3D    & \checkmark   &     & \checkmark   &                   \\ 
\hdashline
Effect of walking variations on complementary filter based inertial data fusion for ankle angle measurement~\cite{Meng2019}                                                                                                 & SF           &     & EST       &     & \checkmark  & \checkmark  &    &                        & 3D    & \checkmark   &     & \checkmark   &                   \\ 
\hdashline
Efficiency of deep neural networks for joint angle modeling in digital gait assessment~\cite{ConteAlcaraz2021}                                                                                                                      & ML          & \checkmark   & ML learns & \checkmark   & \checkmark  & \checkmark  & \checkmark  & \checkmark                      & 2D    & \checkmark   &     & \checkmark   &                   \\ 
\hdashline
Estimate of lower trunk angles in pathological gaits using gyroscope data~\cite{Grimpampi2013}                                                                                                                                  & SF         &     &           & \checkmark   & \checkmark  &    &    &                        & 3D    & \checkmark   &     &     & \checkmark                 \\ 
\hdashline
Estimating Lower Body Kinematics Using a Lie Group Constrained Extended Kalman Filter and Reduced IMU Count~\cite{Sy2021}                                                                                                 & SF          & \checkmark   & \checkmark         & \checkmark   & \checkmark  & \checkmark  & \checkmark  &                        & 3D    & \checkmark   &     &     & \checkmark                 \\ 
\hdashline
Estimating lower extremity running gait kinematics with a single accelerometer: A deep learning approach~\cite{Gholami2020}                                                                                                    & ML          & \checkmark   &           & \checkmark   &    & \checkmark  &    & \checkmark                      & 3D    & \checkmark   &     & \checkmark   &                   \\ 
\hdashline
Estimating Lower Limb Kinematics Using a Reduced Wearable Sensor Count~\cite{Sy2021b}                                                                                                                                      & SF          & \checkmark   & \checkmark         & \checkmark   & \checkmark  & \checkmark  & \checkmark  &                        & 3D    & \checkmark   & \checkmark   & \checkmark   &                   \\ 
\hdashline
Estimation and Observability Analysis of Human Motion on Lie Groups~\cite{Joukov2020}                                                                                                                                         & SF          &     &           & \checkmark   & \checkmark  & \checkmark  &    & & 3D    & \checkmark   &     & \checkmark   &                   \\ 
\hdashline
Estimation of gait kinematics and kinetics from inertial sensor data using optimal control of musculoskeletal models~\cite{Dorschky2019}                                                                                        & OA          &     & \checkmark         & \checkmark   & \checkmark  & \checkmark  &    &                        & 2D    & \checkmark   &     & \checkmark   & \checkmark                 \\ 
\hdashline
Estimation of Gait Mechanics Based on Simulated and Measured IMU Data Using an Artificial Neural Network~\cite{Mundt2020}                                                                                                    & ML           & \checkmark   &           & \checkmark   & \checkmark  & \checkmark  &    & \checkmark                      & 3D    & \checkmark   &     & \checkmark   &                   \\ 
\hdashline
Estimation of kinematics from inertial measurement units using a combined deep learning and optimization framework~\cite{Rapp2021}                                                                                          & ML     & \checkmark   & ML learns & \checkmark   & \checkmark  & \checkmark  &    & \checkmark                      & 3D    & \checkmark   &     & \checkmark   &                   \\ 
\hdashline
Estimation of knee joint angle during gait cycle using inertial measurement unit sensors: a method of sensor-to-clinical bone calibration on the lower limb skeletal model~\cite{Wang2022} & OA          & \checkmark           &              &                  & \checkmark           & \checkmark           &             &             & 3D           & \checkmark            &              & \checkmark            & \checkmark            \\ 
\hdashline
Estimation of lower limb joint angles during walking using extended kalman filtering~\cite{Young2010}                                                                                                                        & SF          &     &           & \checkmark   & \checkmark  & \checkmark  &    &                        & 3D    & \checkmark   &     & \checkmark   &                   \\ 
\hdashline
Estimation of the continuous walking angle of knee and ankle (Talocrural joint, subtalar joint) of a lower‐limb exoskeleton robot using a neural network~\cite{Lee2021}                                                    & ML          & \checkmark   & ML learns & \checkmark   & \checkmark  & \checkmark  &    & \checkmark                      & 2D    & \checkmark   &     & \checkmark   &                   \\ 
\hdashline
Estimation of the knee flexion-extension angle during dynamic sport motions using body-worn inertial sensors~\cite{Jakob2013}                                                                                                & SF          &     &           & \checkmark   & \checkmark  & \checkmark  &    &                        & 2D    & \checkmark   &     & \checkmark   &                   \\ 
\hdashline
Evaluation of wearable gyroscope and accelerometer sensor (PocketIMU2) during walking and sit-to-stand motions~\cite{An2012}                                                                                              & OA          &     &           &     & \checkmark  &    &    &                        & 3D    & \checkmark   &     & \checkmark   &                   \\ 
\hdashline
Feasibility of a Wearable, Sensor-based Motion Tracking System~\cite{Mohammadzadeh2015}                                                                                                                                              & OA          &     &           &     & \checkmark  &    &    &                        & 3D    & \checkmark   &     & \checkmark   &                   \\ 
\hdashline
Feasibility study of inertial sensor-based joint moment estimation method during human movements: A test of multi-link modeling of the trunk segment~\cite{Watanabe2016}                                                        & OA          & \checkmark   & \checkmark         &     & \checkmark  &    &    &                        & 2D    & \checkmark   &     & \checkmark   &                   \\ 
\hdashline
Gait posture estimation using wearable acceleration and gyro sensors~\cite{Takeda2009}                                                                                                                                        & OA          & \checkmark   & \checkmark         & \checkmark   & \checkmark  & \checkmark  &    &                        & 2D    & \checkmark   &     & \checkmark   &                   \\ 
\hdashline
Geometrical kinematic modeling on human motion using method of multi-sensor fusion~\cite{Xu2018}                                                                                                                          & SF           & \checkmark   & \checkmark         &     & \checkmark  & \checkmark  &    &                        & 3D    &     & \checkmark   &     & \checkmark                 \\ 
\hdashline
Grasp pose estimation in human-robot manipulation tasks using wearable motion sensors~\cite{Cehajic2015}                                                                                                                       & OA           &     &           &     & \checkmark  & \checkmark  &    &                        & 3D    & \checkmark   & \checkmark   & \checkmark   &                   \\ 
\hdashline
Hand Motion Measurement using Inertial Sensor System and Accurate Improvement by Extended Kalman Filter~\cite{Kitano2019}                                                                                                     & SF          & \checkmark   & \checkmark         &     & \checkmark  & \checkmark  &    &                        & 3D    & \checkmark   & \checkmark   & \checkmark   &                   \\ 
\hdashline
Human Arm Motion Tracking by Inertial/Magnetic Sensors Using Unscented Kalman Filter and Relative Motion Constraint~\cite{Atrsaei2018}                                                                                         & SF          & \checkmark   & \checkmark         &     & \checkmark  & \checkmark  & \checkmark  &                        & 3D    &     & \checkmark   & \checkmark   &                   \\ 
\hdashline
Human body model based inertial measurement of sit-to-stand motion kinematics~\cite{Music2008}                                                                                                                               & SF          & \checkmark   & \checkmark         &     & \checkmark  & \checkmark  &    &                        & 2D    & \checkmark   &     &     & \checkmark                 \\ 
\hdashline
Human Motion Kinematics Assessment Using Wearable Sensors~\cite{Slajpah2014}                                                                                                                                                   & SF          & \checkmark   & \checkmark         &     & \checkmark  & \checkmark  & \checkmark  &                        & 3D    & \checkmark   &     & \checkmark   &                   \\ 
\hdashline
Human motion tracking based on complementary Kalman filter~\cite{Wang2017}                                                                                                                                                  & SF          &     &           &     & \checkmark  & \checkmark  & \checkmark  &                        & 3D    & \checkmark   &     &     & \checkmark                 \\ 
\hdashline
Human pose recovery for rehabilitation using ambulatory sensors~\cite{Lin2013}                                                                                                                                             & SF          & \checkmark   & \checkmark         &     & \checkmark  & \checkmark  &    &                        & 3D    & \checkmark   &     &     & \checkmark                 \\ 
\hdashline
Human pose recovery using wireless inertial measurement units~\cite{Lin2012}                                                                                                                                               & SF          & \checkmark   & \checkmark         &     & \checkmark  & \checkmark  &    &                        & 3D    & \checkmark   & \checkmark   & \checkmark   &                   \\ 
\hdashline
IMU-Based joint angle measurement for gait analysis~\cite{Seel2014}                                                                                                                                                         & OA        & \checkmark   & EST       &     & \checkmark  & \checkmark  &    &                        & 3D    & \checkmark   &     & \checkmark   &                   \\ 
\hdashline
Inertia-based angle measurement unit for gait assistive device~\cite{Das2014}                                                                                                                                              & OA          &     &           &     &    & \checkmark  &    &                        & 2D    & \checkmark   &     & \checkmark   &                   \\ 
\hdashline
Inertial Motion Capture Using Adaptive Sensor Fusion and Joint Angle Drift Correction~\cite{Butt2019}                                                                                                                       & SF          & \checkmark   &           & \checkmark   & \checkmark  & \checkmark  & \checkmark  &                        & 3D    & \checkmark   &     &     & \checkmark                 \\ 
\hdashline
Inertial sensing in ambulatory back load estimation~\cite{Baten1996}                                                                                                                                                         & OA          &     &           & \checkmark   & \checkmark  & \checkmark  &    &                        & 3D    & \checkmark   &     &     & \checkmark                 \\ 
\hdashline
Inertial sensor-based knee flexion/extension angle estimation~\cite{Cooper2009}                                                                                                                                               & SF           & \checkmark   &           &     & \checkmark  & \checkmark  &    &                        & 3D    & \checkmark   &     & \checkmark   &                   \\ 
\hdashline
Joint angle estimation with accelerometers for dynamic postural analysis~\cite{Ma2015}                                                                                                                                   & OA          & \checkmark   & \checkmark         &     &    & \checkmark  &    &                        & 2D    & \checkmark   &     & \checkmark   &                   \\ 
\hdashline
Joint Inertial Sensor Orientation Drift Reduction for Highly Dynamic Movements~\cite{Fasel2018}                                                                                                                              & OA           & \checkmark   & EST       & \checkmark   & \checkmark  & \checkmark  &    &                        & 3D    & \checkmark   &     & \checkmark   &                   \\ 
\hdashline
Knee angle estimation based on IMU data and artificial neural networks~\cite{Bennett2013}                                                                                                                                      & ML          & \checkmark   &           & \checkmark   & \checkmark  & \checkmark  &    & \checkmark                      & 3D    & \checkmark   &     & \checkmark   &                   \\ 
\hdashline
Knee joint angle measuring portable embedded system based on inertial measurement units for gait analysis~\cite{Mayorca-Torres2020}                                                                                                   & SF           &     &           &     & \checkmark  & \checkmark  &    &                        & 2D    & \checkmark   &     & \checkmark   &                   \\ 
\hdashline
Lower body kinematics estimation from wearable sensors for walking and running: A deep learning approach~\cite{Hernandez2021}                                                                                                   & ML         & \checkmark   & ML learns & \checkmark   & \checkmark  & \checkmark  &    & \checkmark                      & 3D    & \checkmark   &     & \checkmark   &                   \\ 
\hdashline
Lower Extremity Angle Measurement with Accelerometers—Error and Sensitivity~\cite{Willemsen1991}                                                                                                                                 & OA         & \checkmark   & \checkmark         &     &    & \checkmark  &    &                        & 2D    & \checkmark   &     & \checkmark   &                   \\ 
\hdashline
Magneto: Joint angle analysis using an electromagnet-based sensing method~\cite{Watson20211}                                                                                                                                   & OA         &     &           &     &    &    & \checkmark  &                        & 3D    & \checkmark   & \checkmark   & \checkmark   &                   \\ 
\hdashline
Magnetometer robust deep human pose regression with uncertainty prediction using sparse body worn magnetic inertial measurement units~\cite{Butt2021}                                                                       & ML         & \checkmark   & ML learns & \checkmark   & \checkmark  &    & \checkmark  & \checkmark                      & 3D    & \checkmark   &     & \checkmark   &                   \\ 
\hdashline
Monitoring of Hip and Knee Joint Angles Using a Single Inertial Measurement Unit during Lower Limb Rehabilitation~\cite{Bonnet2016}                                                                                           & OA           & \checkmark   & \checkmark         & \checkmark   & \checkmark  & \checkmark  &    &                        & 3D    & \checkmark   &     & \checkmark   &                   \\ 
\hdashline
Multi-Inertial Sensor-Based Arm 3D Motion Tracking Using Elman Neural Network~\cite{Yang2022}                                                                                             & ML          &             &              &                  &  \checkmark          &  \checkmark           &             &  \checkmark           & 3D           &  \checkmark            &  \checkmark            &  \checkmark            &               \\ 
\hdashline
Nonlinear optimization for drift removal in estimation of gait kinematics based on accelerometers~\cite{Djuric-Jovicic2012}                                                                                                           & OA        & \checkmark   & \checkmark         & \checkmark   &    & \checkmark  &    &                        & 2D    & \checkmark   & \checkmark   &     & \checkmark                 \\ 
\hdashline
Novel approach to ambulatory assessment of human segmental orientation on a wearable sensor system~\cite{Liu2009c}                                                                                                          & OA         & \checkmark   & \checkmark         & \checkmark   &    & \checkmark  &    &                        & 2D    & \checkmark   &     & \checkmark   &                   \\ 
\hdashline
Online tracking of the lower body joint angles using IMUs for gait rehabilitation~\cite{Joukov2014}                                                                                                                           & SF          & \checkmark   & \checkmark         &     & \checkmark  & \checkmark  &    &                        & 3D    & \checkmark   & \checkmark   & \checkmark   &                   \\ 
\hdashline
Optimization of Inertial Sensor-Based Motion Capturing for Magnetically Distorted Field Applications~\cite{Schiefer2014}                                                                                                        & OA    &     &           & \checkmark   & \checkmark  & \checkmark  & \checkmark  &                        & 3D    & \checkmark   &     &     & \checkmark                 \\ 
\hdashline
Physical-Sensor and Virtual-Sensor Based Method for Estimation of Lower Limb Gait Posture Using Accelerometers and Gyroscopes~\cite{Liu2010}                                                                               & OA           & \checkmark   & \checkmark         &     & \checkmark  & \checkmark  &    &                        & 3D    & \checkmark   & \checkmark   & \checkmark   &                   \\ 
\hdashline
Pose estimation by extended Kalman filter using noise covariance matrices based on sensor output~\cite{Saito2020}                                                                                                            & SF          &     &           & \checkmark   & \checkmark  & \checkmark  & \checkmark  &                        & 3D    & \checkmark   &     & \checkmark   &                   \\ 
\hdashline
Predicting Knee Joint Kinematics from Wearable Sensor Data in People with Knee Osteoarthritis and Clinical Considerations for Future Machine Learning Models~\cite{Tan2022}               & ML          &             &              &                  & \checkmark           & \checkmark          &             &\checkmark           & 2D           &\checkmark            &              & \checkmark            &               \\ 
\hdashline
Predicting lower limb joint kinematics using wearable motion sensors~\cite{Findlow2008}                                                                                                                                        & ML         & \checkmark   &           & \checkmark   & \checkmark  & \checkmark  &    & \checkmark                      & 2D    & \checkmark   &     & \checkmark   &                   \\ 
\hdashline
Prediction of lower limb joint angles and moments during gait using artificial neural networks~\cite{Mundt2020b}                                                                                                              & ML           & \checkmark   &           & \checkmark   & \checkmark  & \checkmark  &    & \checkmark                      & 3D    & \checkmark   &     & \checkmark   &                   \\ 
\hdashline
Prediction of lower limb kinetics and kinematics during walking by a single IMU on the lower back using machine learning~\cite{Lim2020}                                                                                    & ML          & \checkmark   & \checkmark         & \checkmark   &    & \checkmark  &    &                        & 2D    & \checkmark   & \checkmark   &     & \checkmark                 \\ 
\hdashline
Quasi-real time estimation of angular kinematics using single-axis accelerometers~\cite{Caroselli2013}                                                                                                                           & SF          & \checkmark   & \checkmark         &     & \checkmark  & \checkmark  &    &                        & 2D    & \checkmark   &     & \checkmark   &                   \\ 
\hdashline
Real-time estimate of body kinematics during a planar squat task using a single inertial measurement unit~\cite{Bonnet2013a}                                                                                                   & SF         & \checkmark   & \checkmark         &     & \checkmark  & \checkmark  &    &                        & 2D    & \checkmark   &     & \checkmark   &                   \\ 
\hdashline
Real-Time Human Motion Capture Based on Wearable Inertial Sensor Networks~\cite{Li2022}                                                                                                  & SF          &             &  \checkmark            &  \checkmark                &  \checkmark           &  \checkmark           &  \checkmark           &             & 3D           &  \checkmark            &  \checkmark            &  \checkmark            &               \\
\hdashline
Reconstructing an accelerometer-based pelvis segment for three-dimensional kinematic analyses during laboratory simulated tasks with obstructed line-of-sight~\cite{Zehr2021}                                               & OA          &     & \checkmark         &     &    & \checkmark  &    &                        & 2D    & \checkmark   &     &     & \checkmark                 \\ 
\hdashline
Reconstruction of angular kinematics from wrist-worn inertial sensor data for smart home healthcare~\cite{Villeneuve2017}                                                                                                         & SF       & \checkmark   & \checkmark         &     & \checkmark  & \checkmark  &    &                        & 2D    & \checkmark   &     &     & \checkmark                 \\ 
\hdashline
Reducing drifts in the inertial measurements of wrist and elbow positions~\cite{Zhou2010}                                                                                                                                   & SF           & \checkmark   & \checkmark         &     & \checkmark  & \checkmark  &    &                        & 3D    &     & \checkmark   & \checkmark   &                   \\ 
\hdashline
Research on Human Motion Monitoring Method Based on Multi-Joint Constraint Filter Model~\cite{Yang2021}                                                                                                                     & SF           & \checkmark   & \checkmark         & \checkmark   & \checkmark  & \checkmark  & \checkmark  &                        & 3D    & \checkmark   &     & \checkmark   & \checkmark                 \\ 
\hdashline
Rhythmic EKF for pose estimation during gait~\cite{Joukov2015}                                                                                                                                                                & SF          & \checkmark   & \checkmark         & \checkmark   & \checkmark  & \checkmark  &    &                        & 3D    & \checkmark   &     & \checkmark   &                   \\ 
\hdashline
Rhythmic Extended Kalman Filter for Gait Rehabilitation Motion Estimation and Segmentation~\cite{Joukov2018}                                                                                                                  & SF          & \checkmark   & \checkmark         & \checkmark   & \checkmark  & \checkmark  &    &                        & 3D    & \checkmark   &     & \checkmark   &                   \\ 
\hdashline
Rigid body motion capturing by means of a wearable inertial and magnetic MEMS sensor assembly-from reconstitution of the posture toward dead reckoning: An application in bio-logging~\cite{Fourati2017}                       & SF           &     &           &     & \checkmark  & \checkmark  & \checkmark  &                        & 3D    & \checkmark   & \checkmark   &     & \checkmark                 \\ 
\hdashline
Robust and Accurate Capture of Human Joint Pose Using an Inertial Sensor~\cite{Pathirana2018}                                                                                                                                    & SF          &     &           & \checkmark   & \checkmark  & \checkmark  & \checkmark  &                        & 3D    & \checkmark   &     &     & \checkmark                 \\ 
\hdashline
Sensorial system for obtaining the angles of the human movement in the coronal and sagittal anatomical planes~\cite{Alvarado2017}                                                                                               & SF           & \checkmark   &           &     & \checkmark  & \checkmark  &    &                        & 2D    & \checkmark   &     & \checkmark   & \checkmark                 \\ 
\hdashline
Shoulder and elbow joint angle estimation for upper limb rehabilitation tasks using low-cost inertial and optical sensors~\cite{Alizadegan2017}                                                                                   & SF           & \checkmark   & \checkmark         &     & \checkmark  & \checkmark  &    &                        & 3D    & \checkmark   &     & \checkmark   &                   \\ 
\hdashline
Shoulder and elbow joint angle tracking with inertial sensors~\cite{ElGohary2012}                                                                                                                                               & SF          & \checkmark   & \checkmark         &     & \checkmark  & \checkmark  &    &                        & 3D    & \checkmark   &     & \checkmark   &                   \\ 
\hdashline
Synergy-based knee angle estimation using kinematics of thigh~\cite{Liang2021}                                                                                                                                               & ML         & \checkmark   & ML learns & \checkmark   & \checkmark  & \checkmark  &    & \checkmark                      & 3D    & \checkmark   &     & \checkmark   &                   \\ 
\hdashline
System for measurement of joint range of motion using inertial sensors~\cite{Sharma2017}                                                                                                                                      & SF           & \checkmark   &           &     & \checkmark  & \checkmark  &    &                        & 2D    & \checkmark   &     & \checkmark   &                   \\ 
\hdashline
The head mouse - Head gaze estimation 'In-the-Wild' with low-cost inertial sensors for BMI use~\cite{Sim2013}                                                                                                              & SF           &     &           &     & \checkmark  & \checkmark  & \checkmark  &                        & 3D    & \checkmark   &     &     & \checkmark                 \\ 
\hdashline
The manumeter: A wearable device for monitoring daily use of the wrist and fingers~\cite{Friedman2014}                                                                                                                          & OA         &     &           &     &    &    & \checkmark  & \checkmark                      & 3D    & \checkmark   &     & \checkmark   &                   \\ 
\hdashline
The online estimation of the joint angle based on the gravity acceleration using the accelerometer and gyroscope in the wireless networks~\cite{Ding2020}                                                                   & OA          &     &           &     & \checkmark  & \checkmark  &    &                        & 2D    & \checkmark   &     & \checkmark   &                   \\ 
\hdashline
The use of synthetic IMU signals in the training of deep learning models significantly improves the accuracy of joint kinematic predictions~\cite{SharifiRenani2021}                                                                 & ML         & \checkmark   & ML learns & \checkmark   & \checkmark  & \checkmark  &    & \checkmark                      & 3D    & \checkmark   &     & \checkmark   &                   \\ 
\hdashline
Three dimensional gait analysis using wearable acceleration and gyro sensors based on quaternion calculations~\cite{Tadano2013}                                                                                               & OA          &     &           & \checkmark   & \checkmark  & \checkmark  &    &                        & 3D    & \checkmark   &     & \checkmark   &                   \\ 
\hdashline
Time coherent full-body poses estimated using only five inertial sensors: Deep versus shallow learning~\cite{Wouda2019}                                                                                                      & ML           & \checkmark   &           & \checkmark   & \checkmark  & \checkmark  & \checkmark  &                        & 3D    & \checkmark   & \checkmark   &     & \checkmark                 \\ 
\hdashline
Towards a portable human gait analysis \& monitoring system~\cite{Kumar2018}                                                                                                                                                   & OA          & \checkmark   & EST       &     & \checkmark  & \checkmark  &    &                        & 2D    & \checkmark   &     & \checkmark   &                   \\ 
\hdashline
Towards kinematically constrained real time human pose estimation using sparse IMUs~\cite{Nagaraj2021}                                                                                                                         & ML         & \checkmark   & ML learns & \checkmark   & \checkmark  & \checkmark  & \checkmark  &                        & 3D    & \checkmark   &     & \checkmark   &                   \\ 
\hdashline
Training Data Selection and Optimal Sensor Placement for Deep-Learning-Based Sparse Inertial Sensor Human Posture Reconstruction~\cite{Zheng2021}                                                                           & ML         & \checkmark   & ML learns & \checkmark   & -  & \checkmark  & -  & \checkmark                      & 3D    & \checkmark   &     &     & \checkmark                 \\ 
\hdashline
Ubiquitous human upper-limb motion estimation using wearable sensors~\cite{Zhang2011b}                                                                                                                                        & SF          & \checkmark   &           &     & \checkmark  & \checkmark  & \checkmark  &                        & 3D    & \checkmark   &     & \checkmark   &                   \\ 
\hdashline
Unrestrained measurement of arm motion based on a wearable wireless sensor network~\cite{Lee2010}                                                                                                                          & OA          &     &           & \checkmark   &    & \checkmark  &    &                        & 2D    & \checkmark   &     & \checkmark   &                   \\ 
\hdashline
Upper limb joint angle measurement in occupational health~\cite{Alvarez2016}                                                                                                                                                   & OA           & \checkmark   & \checkmark         & \checkmark   & \checkmark  & \checkmark  & \checkmark  &                        & 3D    & \checkmark   &     & \checkmark   &                   \\ 
\hdashline
Upper limb joint angle tracking with inertial sensors~\cite{ElGohary2011}                                                                                                                                                       & SF          & \checkmark   & \checkmark         &     & \checkmark  & \checkmark  &    &                        & 3D    & \checkmark   &     & \checkmark   &                   \\ 
\hdashline
Use of multiple wearable inertial sensors in upper limb motion tracking~\cite{Zhou2008}                                                                                                                                     & OA          & \checkmark   & \checkmark         &     & \checkmark  & \checkmark  &    &                        & 3D    & \checkmark   & \checkmark   & \checkmark   &                   \\ 
\hdashline
Validation of a low-cost inertial exercise tracker~\cite{Salehi2020}                                                                                                                                                          & SF          & \checkmark   & EST       &  \checkmark & \checkmark  & \checkmark  &    &                        & 3D    & \checkmark   & \checkmark   & \checkmark   &                   \\ 
\hdashline
Visual and quantitative analysis of lower limb 3D gait posture using accelerometers and magnetometers~\cite{Liu2010b}                                                                                                       & OA           & \checkmark   & \checkmark         & -   &    & \checkmark  & \checkmark  &                        & 3D    & \checkmark   &     & \checkmark   &                   \\ 
\hdashline
Visual estimation of lower limb motion using physical and virtual sensors~\cite{Liu2010c}                                                                                                                                   & OA          & \checkmark   & \checkmark         & -   & \checkmark  & \checkmark  & \checkmark  &                        & 3D    & \checkmark   &     & \checkmark   &                   \\ 
\hdashline
Wearable inertial sensor system towards daily human kinematic gait analysis: Benchmarking analysis to MVN BIOMECH~\cite{Figueiredo2020}                                                                                           & ML, SF        & \checkmark   &           & \checkmark   & \checkmark  & \checkmark  &    & \checkmark                      & 2D    & \checkmark   &     & \checkmark &  \checkmark\\
\hline
\end{longtable}
\end{landscape}
\twocolumn

\onecolumn
\begin{landscape}
\begin{longtable}{ p{8.2cm}|p{1.8cm} p{2.8cm}|p{.5cm} p{2.1cm}|p{1.63cm}: p{1.57cm} p{1.6cm} } 
\caption{Relevant details about the monitored anatomic unit, the number of subjects of study in the selected studies the validation system and the metrics for evaluation. LG: lower-limb group, UG: upper-limb group, NS: number of subjects for the analysis, DSS: presence of subjects with diseases, VS: validation sensory system, M1 and M2: metrics provided for the validation of the proposals. }\label{tab:part+validation+subjects} 
\\
\hline
\multirow{2}{*}{\textbf{Title}}   & \multicolumn{2}{c|}{\textbf{Monitored anatomic unit}}                                                                               & \multicolumn{2}{c|}{\textbf{Subjects of study}}    & \multicolumn{3}{c}{\textbf{Validation approach}}                                     
\\
      & \textbf{LG}                        & \textbf{UG}                                                                               & \textbf{NS} & \textbf{DSS}                                    & \textbf{VS}                                                         & \textbf{M1}                                                                                  & \textbf{M2}                                       \\ 
\endfirsthead
\hline
\multirow{2}{*}{\textbf{Title}}   & \multicolumn{2}{c|}{\textbf{Monitored anatomic unit}}                                                                               & \multicolumn{2}{c|}{\textbf{Subjects of study}}    & \multicolumn{3}{c}{\textbf{Validation approach}}                                     
\\
      & \textbf{LG}                        & \textbf{UG}                                                                               & \textbf{NS} & \textbf{DSS}                                    & \textbf{VS}                                                         & \textbf{M1}                                                                                  & \textbf{M2}                                       \\ 
      \endhead
\hdashline
A basic study on variable-gain Kalman filter based on angle error calculated from acceleration signals for lower limb angle measurement with inertial sensors~\cite{Teruyama2013}                                            & Leg Feet                  &                                                                                  & 3                       &                                        & Optical 3D                                                 & RMSE                                                                                &                                          \\ 
\hdashline
A Comparison of Three Neural Network Approaches for Estimating Joint Angles and Moments from Inertial Measurement Units~\cite{Mundt2021}                                                                                     & Hip/knee/ankle            &                                                                                  & 116                     &                                        & Optical 3D, Force plates                                   & Difference in nRMSE between methods                                                 &                                          \\ 
\hdashline
A fast quaternion-based orientation optimizer via virtual rotation for human motion tracking~\cite{Lee2009}                                                                                                                  &                           & Arm                                                                              & 1                       &                                        & Optical 3D                                                 & RMSE                                                                                &                                          \\ 
\hdashline
A Full-State Robust Extended Kalman Filter for Orientation Tracking during Long-Duration Dynamic Tasks Using Magnetic and Inertial Measurement Units~\cite{Nazarahari2021}                                                   & Leg Feet                  &                                                                                  & 9                       &                                        & Optical 3D                                                 & RMSE                                                                                &                                          \\ 
\hdashline
A Human Motion Tracking Algorithm Using Adaptive EKF Based on Markov Chain~\cite{Kang2016}                                                                                                                                   &                           & Arm                                                                              & 1                       &                                        & Optical 3D                                                 & RMSE                                                                                &                                          \\ 
\hdashline
A method for gait analysis in a daily living environment by body-mounted instruments~\cite{Ohtaki2001}                                                                                                                       & Hip/knee/ankle            &                                                                                  & 6                       &                                        & Optical 3D, Force plates                                   & RMSE                                                                                & -                                        \\ 
\hdashline
A Method for Lower Back Motion Assessment Using Wearable 6D Inertial Sensors~\cite{Molnar2018}                                                                                                                               &                           & Back/trunk/torso                                                                 & 8                       &                                        & Optical 3D                                                 & RMSE                                                                                & -                                        \\ 
\hdashline
A Miniature Sensor System for Precise Hand Position Monitoring~\cite{Abbasi-Kesbi2018}                                                                                                                                       &                           & Arm joints: shoulders/elbow/forearm twist Hand/wrist/fingers                     & 6                       &                                        & Optical 3D                                                 & RMSE                                                                                & -                                        \\ 
\hdashline
A new approach to accurate measurement of uniaxial joint angles based on a combination of accelerometers and gyroscopes~\cite{Dejnabadi2005}                                                                                 & Hip/knee/ankle            &                                                                                  & 8                       &                                        & Visual 2D                                                  & RMSE                                                                                & Corr. coeff.                               \\ 
\hdashline
A new quaternion-based kalman filter for human body motion tracking using the second estimator of the optimal quaternion algorithm and the joint angle constraint method with inertial and magnetic sensors~\cite{Duan20201} &                           & Arm                                                                              & 1                       &                                        & Optical 3D                                                 & RMSE                                                                                &                                          \\ 
\hdashline
A nonlinear dynamics-based estimator for functional electrical stimulation: Preliminary results from lower-leg extension experiments~\cite{Allen2017}                                                                        & Hip/knee/ankle            &                                                                                  & 3                       &                                        & Other                                                    & RMSE                                                                                & -                                        \\ 
\hdashline
A novel 7 degrees of freedom model for upper limb kinematic reconstruction based on wearable sensors~\cite{Peppoloni2013}                                                                                                    &                           & Head/neck/scapula Back/trunk/torso Arm Arm joints: shoulders/elbow/forearm twist & 1                       &                                        & Optical 3D                                                 & RMSE                                                                                & -                                        \\ 
\hdashline
A Novel Application of Flexible Inertial Sensors for Ambulatory Measurement of Gait Kinematics~\cite{Yin2021}                                                                                                                & Hip/knee/ankle            &                                                                                  & 3                       &                                        & Optical 3D                                                 & RMSE                                                                                &                                          \\ 
\hdashline
A novel approach to motion tracking with wearable sensors based on Probabilistic Graphical Models~\cite{Ruffaldi2014}                                                                                                        &                           & Arm joints: shoulders/elbow/forearm twist                                        & 1                       &                                        & Optical 3D                                                 & RMSE                                                                                &                                          \\ 
\hdashline
A novel complimentary filter for tracking hip angles during cycling using wireless inertial sensors and dynamic acceleration estimation~\cite{Cockcroft2014}                                                                 & Hip/knee/ankle            &                                                                                  & 1                       &                                        & Optical 3D                                                 & MAE                                                                                 & -                                        \\ 
\hdashline
A novel data glove for fingers motion capture using inertial and magnetic measurement units~\cite{Fang2016}                                                                                                                  &                           & Hand/wrist/fingers                                                               & 1                       &                                        & Visual 2D, Reconstruction                                  & RMSE                                                                                & -                                        \\ 
\hdashline
A novel hierarchical information fusion method for three-dimensional upper limb motion estimation~\cite{Zhang2011}                                                                                                           &                           & Arm Arm joints: shoulders/elbow/forearm twist                                    & 4                       &                                        & Optical 3D                                                 & RMSE                                                                                & -                                        \\ 
\hdashline
A Novel Kalman Filter for Human Motion Tracking With an Inertial-Based Dynamic Inclinometer~\cite{Ligorio2015}                                                                                                               &                           & Arm                                                                              & 10                      &                                        & Optical 3D                                                 & RMSE                                                                                & -                                        \\ 
\hdashline
A novel method for estimating knee angle using two leg-mounted gyroscopes for continuous monitoring with mobile health devices~\cite{Allseits2018}                                                                           & Hip/knee/ankle            &                                                                                  & 5                       &                                        & Optical 3D                                                 & LAM & Avg. error                                                                                           \\ 
\hdashline
A novel method of using accelerometry for upper limb FES control~\cite{Sun2016}                                                                                                                                              &                           & Arm                                                                              & 2                       & 2 post-stroke                          & Optical 3D                                                 & MAE                                                                                 &                                          \\ 
\hdashline
A novel sensor-based assessment of lower limb spasticity in children with cerebral palsy~\cite{Choi2018}                                                                                                                     & Hip/knee/ankle            &                                                                                  & 33                      & 28 patients cerebral palsy             & Optical 3D                                                 & RMSE                                                                                & Test-retest  inter-rater reliabilities  \\ 
\hdashline
A patient-centric sensory system for in-home rehabilitation~\cite{Abbasi-Kesbi2017}                                                                                                                                          &                           & Arm joints: shoulders/elbow/forearm twist Hand/wrist/fingers                     & 1                       &                                        & Other                   & RMSE                                                                                &                                          \\ 
\hdashline
A preliminary test of measurement of joint angles and stride length with wireless inertial sensors for wearable gait evaluation system~\cite{watanabe2011}                                                                   & Hip/knee/ankle Feet       &                                                                                  & 3                       &                                        & Optical 3D                                                 & RMSE                                                                                & Corr. coeff.                               \\ 
\hdashline
A study of gait analysis with a smartphone for measurement of hip joint angle~\cite{Niijima2014}                                                                                                                             & Hip/knee/ankle            &                                                                                  & 14                      &                                        & Optical 3D                                                 & RMSE                                                                                &             \\ 
\hdashline
A Wearable Human Motion Tracking Device Using Micro Flow Sensor Incorporating a Micro Accelerometer~\cite{Liu2020}                                                                                                           &                           & Arm                                                                              & 1                       &                                        & IMU                                  & RMSE                                                                                &                                          \\ 
\hdashline
A Wearable Magnetometer-Free Motion Capture System: Innovative Solutions for Real-World Applications~\cite{Ligorio2020}                                                                                                      &                           & Arm joints: shoulders/elbow/forearm twist                                        & 10                      &                                        & Optical 3D                                                 & RMSE                                                                                &                                          \\ 
\hdashline
Accuracy Improvement on the Measurement of Human-Joint Angles~\cite{Meng2016}                                                                                                                                                &                           & Arm joints: shoulders/elbow/forearm twist                                        & 15                      &                                        & Optical 3D                                                 & RMSE                                                                                &                                          \\ 
\hdashline
Accuracy of a custom physical activity and knee angle measurement sensor system for patients with neuromuscular disorders and gait abnormalities~\cite{Feldhege2015}                                                         & Hip/knee/ankle            &                                                                                  & 10                      &                                        & Goniometer                                         & RMSE                                                                                &                                          \\ 
\hdashline
Adaptive Gain Regulation of Sensor Fusion Algorithms for Orientation Estimation with Magnetic and Inertial Measurement Units~\cite{Nazarahari2021b}                                                                          & Leg Feet                  &                                                                                  & 9                       &                                        & Optical 3D                                                 & RMSE                                                                                &                                          \\ 
\hdashline
Alignment-Free, Self-Calibrating Elbow Angles Measurement Using Inertial Sensors~\cite{Muller2017}                                                                                                                           &                           & Arm joints: shoulders/elbow/forearm twist                                        & 1                       &                                        & Optical 3D                                                 & RMSE                                                                                &                                          \\ 
\hdashline
Ambulatory estimation of knee-joint kinematics in anatomical coordinate system using accelerometers and magnetometers~\cite{Kun2011}                                                                                         & Hip/knee/ankle            &                                                                                  & 5                       &                                        & Optical 3D                                                 & RMSE                                                                                &                                          \\ 
\hdashline
Ambulatory measurement and analysis of the lower limb 3D posture using wearable sensor system~\cite{Liu2009}                                                                                                                 & Hip/knee/ankle            &                                                                                  & 8                       &                                        & Optical 3D                                                 & RMSE                                                                                &                                          \\ 
\hdashline
An adaptive complementary filter for inertial sensor based data fusion to track upper body motion~\cite{Karunarathne2014}                                                                                                    &                           & Arm joints: shoulders/elbow/forearm twist                                        & 4                       &                                        & Optical 3D                                                 & RMSE                                                                                &                                          \\ 
\hdashline
An auto-calibrating knee flexion-extension axis estimator using principal component analysis with inertial sensors~\cite{McGrath2018}                                                                                        & Hip/knee/ankle            &                                                                                  & 15                      &                                        & Optical 3D                                   & RMSE                                                                                &                                          \\ 
\hdashline
An inertial human upper limb motion tracking method for robot programming by demonstration~\cite{Pellois2022}                                                                                &                                 & Hand/wrist/fingers                                           & 1           &                         & IMU & MAE         &                       \\ 
\hdashline
An inertial sensor system for measurements of tibia angle with applications to knee valgus/varus detection~\cite{Hu2014}                                                                                                     & Leg                       &                                                                                  & 3                       &                                        & Optical 3D                                                 & RMSE                                                                                &                                          \\ 
\hdashline
An instance-based algorithm with auxiliary similarity information for the estimation of gait kinematics from wearable sensors~\cite{Goulermas2008}                                                                           & Hip/knee/ankle            &                                                                                  & 8                       &                                        & Optical 3D                                                 & Absolute deviation
\\ 
\hdashline
An investigation into the accuracy of calculating upper body joint angles using MARG sensors~\cite{Mazomenos2014}                                                                                                            &                           & Back/trunk/torso Arm joints: shoulders/elbow/forearm twist Hand/wrist/fingers    & 1                       &                                        & RBG + depth                                                & RMSD                                                                                &                                          \\ 
\hdashline
An optimized Kalman filter for the estimate of trunk orientation from inertial sensors data during treadmill walking\textbackslash{}cite{Mazza2012}                                                                                                         &                           & Back/trunk/torso                                                                 & 18                      &                                        & Optical 3D                                                 & RMSE                                                                                &                                          \\ 
\hdashline
Analysis of a mobile system to register the kinematic parameters in ankle, knee, and hip based in inertial sensors~\cite{Flores-Morales2016}                                                                                 & Hip/knee/ankle            &                                                                                  & 1                       &                                        & Visual 2D                                                  & Max. error                                                           &                                          \\ 
\hdashline
Analyzing 3D knee kinematics using accelerometers, gyroscopes and magnetometers~\cite{Kawano2007}                                                                                                                            & Hip/knee/ankle            &                                                                                  & 1                       &                                        & Optical 3D                                                 & Avg. error                                                                          &                                          \\ 
\hdashline
Angle measurements during 2D and 3D movements of a rigid body model of lower limb: Comparison between integral-based and quaternion-based methods~\cite{Watanabe2014}                                                        & Leg                       &                                                                                  & 1                       &                                        & Optical 3D                                                 & RMSE                                                                                &                                          \\ 
\hdashline
Artificial neural networks in motion analysis—applications of unsupervised and heuristic feature selection techniques~\cite{Mundt20201}                                                                                      & Hip/knee/ankle            &                                                                                  & 115                     & 24 knee arthroplasty                   & Optical 3D                                                 & RMSE                                                                                & Corr. coeff.                        \\ 
\hdashline
Biomechanical Ambulatory Assessment of 3D Knee Angle Using Novel Inertial Sensor-Based Technique~\cite{Tham2021}                                                                                                             & Hip/knee/ankle            &                                                                                  & 5                       &                                        & Optical 3D                                                 & RMSE                                                                                &                                          \\ 
\hdashline
Calibrated 2D angular kinematics by single-axis accelerometers: From inverted pendulum to N-Link chain~\cite{Bagala2012}                                                                                                     &                           & Back/trunk/torso                                                                 & 1                       &                                        & Optical 3D                                                 & RMSE                                                                                &                                          \\ 
\hdashline
Closed-chain pose estimation from wearable sensors~\cite{Joukov2019}                                                                                                                                                         & Hip/knee/ankle            &                                                                                  & 3                       &                                        & Optical 3D                                                 & Avg. error                                    & RMSE                  \\ 
\hdashline
CNN-Based Estimation of Sagittal Plane Walking and Running Biomechanics From Measured and Simulated Inertial Sensor Data~\cite{Dorschky2020}                                                                                 & Hip/knee/ankle            &                                                                                  & 10                      &                                        & Optical 3D, Force plates                                   & RMSE                                                                                &                                          \\ 
\hdashline
Comparison of angle measurements between integral-based and quaternion-based methods using inertial sensors for gait evaluation~\cite{Watanabe2015}                                                                          & Leg Feet                  &                                                                                  & 3                       &                                        & Optical 3D                                                 & RMSE                                                                                &                                          \\ 
\hdashline
Data-Driven Reconstruction of Human Locomotion Using a Single Smartphone~\cite{Eom2014}                                                                                                                                      &                           & Head/neck/scapula Back/trunk/torso Arm                                           & 1                       &                                        & Optical 3D                                                 & RMSE                                                                                &                                          \\\hdashline
DeepBBWAE-Net: A CNN-RNN Based Deep SuperLearner For Estimating Lower Extremity Sagittal Plane Joint Kinematics Using Shoe-Mounted IMU Sensors In Daily Living~\cite{Hossain2022}            & Hip/knee/ankle                  &                                                              & 10          &                         & Optical 3D                & RMSE        &                       \\ 
 
\hdashline
Deriving kinematic quantities from accelerometer readings for assessment of functional upper limb motions~\cite{Laidig2017}                                                                                                  &                           & Hand/wrist/fingers                                                               & 1                       &                                        & Optical 3D                                                 & RMSE                                                                                &                                          \\ 
\hdashline
Design and validation of an ambulatory inertial system for 3-D measurements of low back movements~\cite{Charry2011}                                                                                                          &                           & Back/trunk/torso                                                                 & 2                       &                                        & Optical 3D                                                 & RMSE                                                                                &                                          \\ 
\hdashline
Detecting absolute human knee angle and angular velocity using accelerometers and rate gyroscopes~\cite{Williamson2001}                                                                                                      & Hip/knee/ankle            &                                                                                  & 2                       & 1 Paraplegic                           & Goniometer                                         & RMSE                                                                                &                                          \\ 
\hdashline
Development of a body joint angle measurement system using IMU sensors~\cite{Bakhshi2011}                                                                                                                                    & Hip/knee/ankle            &                                                                                  & 1                       &                                        & Optical 3D                                                 & Avg. error                                                                       &                                          \\ 
\hdashline
Development of a wearable glove system with multiple sensors for hand kinematics assessment~\cite{Fei2021}                                                                                                                   &                           & Hand/wrist/fingers                                                               & 1                       &                                        &  Goniometer & MAE                                                                          &                                          \\ 
\hdashline
Development of a wearable sensor system for quantitative gait analysis~\cite{Liu2009}                                                                                                                                        & Leg                       &                                                                                  & 10                      &                                        & Optical 3D                                                 & RMSE                                                                                & Corr. coeff.                               \\ 
\hdashline
Digital inclinometer for joint angles measurements with a real-time 3D-animation~\cite{Zabat2015}                                                                                                                            &                           & Head/neck/scapula Arm joints: shoulders/elbow/forearm twist                      & 1                       &                                        & Goniometer                & Max. error                                                                            &                                          \\ 
\hdashline
Drift-Free and Self-Aligned IMU-Based Human Gait Tracking System with Augmented Precision and Robustness~\cite{Chen2020}                                                                                                     & Hip/knee/ankle            &                                                                                  & 1                       &                                        & Optical 3D                                                 & RMSE                                                                                &                                          \\ 
\hdashline
Drift-Free Foot Orientation Estimation in Running Using Wearable IMU~\cite{Falbriard2020}                                                                                                                                    & Feet                      &                                                                                  & 26                      &                                        & Optical 3D                                                 & RMSE                                                                                &                                          \\ 
\hdashline
Drift-Free 3D Orientation and Displacement Estimation for Quasi-Cyclical Movements Using One Inertial Measurement Unit: Application to Running~\cite{Zandbergen2022}                             & Leg                             &                                                              & 4           &                         & Optical 3D                & RMSE        &                       \\ 
\hdashline
Drift-Free Inertial Sensor-Based Joint Kinematics for Long-Term Arbitrary Movements~\cite{Weygers2020}                                                                                                                       & Hip/knee/ankle            &                                                                                  & 11                      &                                        & Optical 3D                                                 & RMSE                                                                                &                                          \\ 
\hdashline
Effect of walking variations on complementary filter based inertial data fusion for ankle angle measurement~\cite{Meng2019}                                                                                                  & Hip/knee/ankle            &                                                                                  & 10                      &                                        & Optical 3D                                                 & RMSE                                                                                &                                          \\ 
\hdashline
Efficiency of deep neural networks for joint angle modeling in digital gait assessment~\cite{ConteAlcaraz2021}                                                                                                               & Hip/knee/ankle            &                                                                                  & 20                      &                                        & IMU                                  & RMSE                                                                                &                                          \\ 
\hdashline
Estimate of lower trunk angles in pathological gaits using gyroscope data~\cite{Grimpampi2013}                                                                                                                               &                           & Back/trunk/torso                                                                 & 27                      & 13 hemiplegia, 14 Parkinson           & Optical 3D                                                 & RMSE                                                                                &                                          \\ 
\hdashline
Estimation of knee joint angle during gait cycle using inertial measurement unit sensors: a method of sensor-to-clinical bone calibration on the lower limb skeletal model~\cite{Wang2022} & Leg Hip/knee/ankle              &                                                              & 8           &                         & Optical 3D                & RMSE        &                       \\ 
\hdashline
Estimating Lower Body Kinematics Using a Lie Group Constrained Extended Kalman Filter and Reduced IMU Count~\cite{Sy2021}                                                                                                    & Leg Hip/knee/ankle        &                                                                                  & 14                      &                                        & Optical 3D                                                 &                                                                    MAE                 &                                          \\ 
\hdashline
Estimating lower extremity running gait kinematics with a single accelerometer: A deep learning approach~\cite{Gholami2020}                                                                                                  & Hip/knee/ankle            &                                                                                  & 10                      &                                        & Optical 3D                                                 & RMSE                                                                                &                                          \\ 
\hdashline
Estimating Lower Limb Kinematics Using a Reduced Wearable Sensor Count~\cite{Sy2021b}                                                                                                                                        & Hip/knee/ankle            &                                                                                  & 9                       &                                        & Optical 3D                                                 & RMSE                                                                                &                                          \\ 
\hdashline
Estimation and Observability Analysis of Human Motion on Lie Groups~\cite{Joukov2020}                                                                                                                                        &                           &                                                                                  & -  &                                        & Optical 3D                                                 & MAE                                                                                 &                                          \\ 
\hdashline
Estimation of gait kinematics and kinetics from inertial sensor data using optimal control of musculoskeletal models~\cite{Dorschky2019}                                                                                     & Pelvis Leg Hip/knee/ankle &                                                                                  & 10                      &                                        & Optical 3D, Force plates                                   & RMSE                                                                                &                                          \\ 
\hdashline
Estimation of Gait Mechanics Based on Simulated and Measured IMU Data Using an Artificial Neural Network~\cite{Mundt2020}                                                                                                    & Hip/knee/ankle            &                                                                                  & 30                      & 24 Knee arthroplasty        & Optical 3D, Force plates                                   & RMSE                                                                                &                                          \\ 
\hdashline
Estimation of kinematics from inertial measurement units using a combined deep learning and optimization framework~\cite{Rapp2021}                                                                                           & Hip/knee/ankle            &                                                                                  & 1000                    & running-related injury & Optical 3D                                                 & RMSE                                                                                &                                          \\ 
\hdashline
Estimation of lower limb joint angles during walking using extended kalman filtering~\cite{Young2010}                                                                                                                        & Leg Hip/knee/ankle        &                                                                                  & 3                       &                                        & Optical 3D                                                 & MAE                                                                         &                                          \\ 
\hdashline
Estimation of the continuous walking angle of knee and ankle (Talocrural joint, subtalar joint) of a lower‐limb exoskeleton robot using a neural network~\cite{Lee2021}                                                      & Hip/knee/ankle            &                                                                                  & 1                       &                                        & Other                                                    & MAE                                                                                 &                                          \\ 
\hdashline
Estimation of the knee flexion-extension angle during dynamic sport motions using body-worn inertial sensors~\cite{Jakob2013}                                                                                                & Leg Hip/knee/ankle        &                                                                                  & 7                       &                                        & Optical 3D                                                 & RMSE                                                                                &                                          \\ 
\hdashline
Evaluation of wearable gyroscope and accelerometer sensor (PocketIMU2) during walking and sit-to-stand motions~\cite{An2012}                                                                                                 & Leg Hip/knee/ankle        &                                                                                  & 1                       &                                        & Optical 3D                                                 & RMSE                                                                                &                                          \\ 
\hdashline
Feasibility of a Wearable, Sensor-based Motion Tracking System~\cite{Mohammadzadeh2015}                                                                                                                                      &                           & Arm Arm joints: shoulders/elbow/forearm twist                                    & 8                       &                                        & Optical 3D                                                 & RMSE                                                                                &                                          \\ 
\hdashline
Feasibility study of inertial sensor-based joint moment estimation method during human movements: A test of multi-link modeling of the trunk segment~\cite{Watanabe2016}                                                     &                           & Back/trunk/torso                                                                 & 3                       &                                        & Optical 3D                                                 & RMSE                                                                                &                                          \\ 
\hdashline
Gait posture estimation using wearable acceleration and gyro sensors~\cite{Takeda2009}                                                                                                                                       & Hip/knee/ankle            &                                                                                  & 3                       &                                        & Optical 3D                                                 & RMSE                                                                                &                                          \\ 
\hdashline
Geometrical kinematic modeling on human motion using method of multi-sensor fusion~\cite{Xu2018}                                                                                                                             & Hip/knee/ankle            &                                                                                  & 1                       &                                        & Other                                              & RMSE                                                                                &                                          \\ 
\hdashline
Grasp pose estimation in human-robot manipulation tasks using wearable motion sensors~\cite{Cehajic2015}                                                                                                                     &                           & Hand/wrist/fingers                                                               & 1                       &                                        & Other                                                      & Absolute errors                                                                     &                                          \\ 
\hdashline
Hand Motion Measurement using Inertial Sensor System and Accurate Improvement by Extended Kalman Filter~\cite{Kitano2019}                                                                                                    &                           & Arm joints: shoulders/elbow/forearm twist Hand/wrist/fingers                     & 1                       &                                        & Optical 3D                                                 & RMSE                                                                                &                                          \\ 
\hdashline
Human Arm Motion Tracking by Inertial/Magnetic Sensors Using Unscented Kalman Filter and Relative Motion Constraint~\cite{Atrsaei2018}                                                                                       &                           & Arm joints: shoulders/elbow/forearm twist Hand/wrist/fingers                     & 1                       &                                        & Optical 3D                                                 & RMSE                                                                                &                                          \\ 
\hdashline
Human body model based inertial measurement of sit-to-stand motion kinematics~\cite{Music2008}                                                                                                                               &                           & Back/trunk/torso                                                                 & 1                       &                                        & Optical 3D                                                 & RMSE                                                                                &                                          \\ 
\hdashline
Human Motion Kinematics Assessment Using Wearable Sensors~\cite{Slajpah2014}                                                                                                                                                 & Hip/knee/ankle            &                                                                                  & 5                       &                                        & Optical 3D, Force plates                                   & Absolute errors                                                                     &                                          \\ 
\hdashline
Human motion tracking based on complementary Kalman filter~\cite{Wang2017}                                                                                                                                                   & Leg                       &                                                                                  & 1                       &                                        & Optical 3D                                                 & RMSE                                                                                &                                          \\ 
\hdashline
Human pose recovery for rehabilitation using ambulatory sensors~\cite{Lin2013}                                                                                                                                               & Leg                       &                                                                                  & 7                       & Joint replacement inpatients           & Other                                                  & RMSE                                                                                &                                          \\ 
\hdashline
Human pose recovery using wireless inertial measurement units~\cite{Lin2012}                                                                                                                                                 & Hip/knee/ankle            &                                                                                  & 20                      &                                        & Optical 3D                                                 & RMSE                                                                                &                                          \\ 
\hdashline
IMU-Based joint angle measurement for gait analysis~\cite{Seel2014}                                                                                                                                                          & Hip/knee/ankle            &                                                                                  & 1                       & Transfemoral amputee                 & Optical 3D                                                 & RMSE                                                                                &                                          \\ 
\hdashline
Inertia-based angle measurement unit for gait assistive device~\cite{Das2014}                                                                                                                                                & Hip/knee/ankle            &                                                                                  & 1                       &                                        & Goniometer                                         &                                                                                     &                                          \\ 
\hdashline
Inertial Motion Capture Using Adaptive Sensor Fusion and Joint Angle Drift Correction~\cite{Butt2019}                                                                                                                        & Hip/knee/ankle            &                                                                                  & 1                       &                                        & Optical 3D                                                 & Error                                                                               &                                          \\ 
\hdashline
Inertial sensing in ambulatory back load estimation~\cite{Baten1996}                                                                                                                                                         &                           & Back/trunk/torso                                                                 & 9                       &                                        & Optical 3D                                                 & \% Error                                                                            &                                          \\ 
\hdashline
Inertial sensor-based knee flexion/extension angle estimation~\cite{Cooper2009}                                                                                                                                              & Hip/knee/ankle            &                                                                                  & 7                       &                                        & Optical 3D                                                 & RMSE                                                                                &                                          \\ 
\hdashline
Joint angle estimation with accelerometers for dynamic postural analysis~\cite{Ma2015}                                                                                                                                       & Hip/knee/ankle            &                                                                                  & 1                       &                                        & Other                                        & RMSE                                                                                &                                          \\ 
\hdashline
Joint Inertial Sensor Orientation Drift Reduction for Highly Dynamic Movements~\cite{Fasel2018}                                                                                                                              &                           & Back/trunk/torso                                                                 & 6                       &                                        & Visual 2D                                                  & MAE                                                                                 &                                          \\ 
\hdashline
Knee angle estimation based on IMU data and artificial neural networks~\cite{Bennett2013}                                                                                                                                    & Hip/knee/ankle            &                                                                                  & 1                       &                                        & Goniometer                                         & RMSE                                                                                &                                          \\ 
\hdashline
Knee joint angle measuring portable embedded system based on inertial measurement units for gait analysis~\cite{Mayorca-Torres2020}                                                                                          & Hip/knee/ankle            &                                                                                  & 12                      &                                        & Visual 2D                                                  & RMSE                                                                                &                                          \\ 
\hdashline
Lower body kinematics estimation from wearable sensors for walking and running: A deep learning approach~\cite{Hernandez2021}                                                                                                & Pelvis Hip/knee/ankle     &                                                                                  & 27                      &                                        & Optical 3D                                                 & MAE                                                                                 &                                          \\ 
\hdashline
Lower Extremity Angle Measurement with Accelerometers—Error and Sensitivity~\cite{Willemsen1991}                                                                                                                             & Hip/knee/ankle            &                                                                                  & 1                       &                                        & Optical 3D                                                 & Avg. errors                                                                         &                                          \\ 
\hdashline
Magneto: Joint angle analysis using an electromagnet-based sensing method~\cite{Watson20211}                                                                                                                                 &                           & Arm joints: shoulders/elbow/forearm twist                                        & 13                      &                                        & Goniometer                                                 & MAE                                                                                 &                                          \\ 
\hdashline
Magnetometer robust deep human pose regression with uncertainty prediction using sparse body worn magnetic inertial measurement units~\cite{Butt2021}                                                                        &                           &                                                                                  &   &                                        & Other                        & RMSE                                                                                &                                          \\ 
\hdashline
Monitoring of Hip and Knee Joint Angles Using a Single Inertial Measurement Unit during Lower Limb Rehabilitation~\cite{Bonnet2016}                                                                                          & Hip/knee/ankle            &                                                                                  & 10                      &                                        & Optical 3D                                                 & RMSD                                                                                &                                          \\ 
\hdashline
Multi-Inertial Sensor-Based Arm 3D Motion Tracking Using Elman Neural Network~\cite{Yang2022}                                                                                              &                                 & Arm joints: shoulders/elbow/forearm twist Hand/wrist/fingers & 1           &                         & Optical 3D                & MAE         &                       \\ 
\hdashline
Nonlinear optimization for drift removal in estimation of gait kinematics based on accelerometers~\cite{Djuric-Jovicic2012}                                                                                                  & Leg                       &                                                                                  & 10                      &                                        & Optical 3D                                                 & RMSE                                                                                &                                          \\ 
\hdashline
Novel approach to ambulatory assessment of human segmental orientation on a wearable sensor system~\cite{Liu2009c}                                                                                                           & Leg                       &                                                                                  & 8                       &                                        & Optical 3D                                                 & RMSE                                                                                &                                          \\ 
\hdashline
Online tracking of the lower body joint angles using IMUs for gait rehabilitation~\cite{Joukov2014}                                                                                                                          & Hip/knee/ankle            &                                                                                  & 5                       &                                        & Optical 3D                                                 & RMSE                                                                                &                                          \\ 
\hdashline
Optimization of Inertial Sensor-Based Motion Capturing for Magnetically Distorted Field Applications~\cite{Schiefer2014}                                                                                                     &                           & Head/neck/scapula Back/trunk/torso Arm                                           & 8                       &                                        & Optical 3D                                                 & RMSE                                                                                &                                          \\ 
\hdashline
Physical-Sensor and Virtual-Sensor Based Method for Estimation of Lower Limb Gait Posture Using Accelerometers and Gyroscopes~\cite{Liu2010}                                                                                 & Leg Hip/knee/ankle        &                                                                                  & 5                       &                                        & Optical 3D                                                 & RMSE                                                                                &                                          \\ 
\hdashline
Pose estimation by extended Kalman filter using noise covariance matrices based on sensor output~\cite{Saito2020}                                                                                                            & Hip/knee/ankle            &                                                                                  & 3                       &                                        & Optical 3D                                                 & RMSE                                                                                &                                          \\ 
\hdashline
Predicting Knee Joint Kinematics from Wearable Sensor Data in People with Knee Osteoarthritis and Clinical Considerations for Future Machine Learning Models~\cite{Tan2022}               & Hip/knee/ankle                  &                                                              & 17          & 17 osteoarthritis       & Optical 3D                & RMSE        &                       \\ 
\hdashline
Predicting lower limb joint kinematics using wearable motion sensors~\cite{Findlow2008}                                                                                                                                      & Hip/knee/ankle            &                                                                                  & 8                       &                                        & Optical 3D                                                 & MAE                                                                                 &                                          \\ 
\hdashline
Prediction of lower limb joint angles and moments during gait using artificial neural networks~\cite{Mundt2020b}                                                                                                             & Hip/knee/ankle            &                                                                                  & 12                      &                                        & Optical 3D                                                 & RMSE                                                                                &                                          \\ 
\hdashline
Prediction of lower limb kinetics and kinematics during walking by a single IMU on the lower back using machine learning~\cite{Lim2020}                                                                                      & Leg Feet                  &                                                                                  & 7                       &                                        & Optical 3D, Force plates                                   & \%RMSE                                                                              &                                          \\ 
\hdashline
Quasi-real time estimation of angular kinematics using single-axis accelerometers~\cite{Caroselli2013}                                                                                                                       & Hip/knee/ankle            &                                                                                  & 1                       &                                        & Optical 3D                                                 & RMSE                                                                                &                                          \\ 
\hdashline
Real-time estimate of body kinematics during a planar squat task using a single inertial measurement unit~\cite{Bonnet2013a}                                                                                                 & Hip/knee/ankle            &                                                                                  & 8                       &                                        & Optical 3D                                                 & RMSE                                                                                &                                          \\ 
\hdashline
Real-Time Human Motion Capture Based on Wearable Inertial Sensor Networks~\cite{Li2022}                                                                                                  & Hip/knee/ankle Feet             &                                                              & 1           &                         & Optical 3D                & RMSE        &                       \\ 
\hdashline
Reconstructing an accelerometer-based pelvis segment for three-dimensional kinematic analyses during laboratory simulated tasks with obstructed line-of-sight~\cite{Zehr2021}                                                & Pelvis                    &                                                                                  & 1                       &                                        & Optical 3D                                                 & RMSE                                                                                &                                          \\ 
\hdashline
Reconstruction of angular kinematics from wrist-worn inertial sensor data for smart home healthcare~\cite{Villeneuve2017}                                                                                                    &                           & Arm                                                                              & 2                       &                                        & IMU                                  & MAE                                                                                 &                                          \\ 
\hdashline
Reducing drifts in the inertial measurements of wrist and elbow positions~\cite{Zhou2010}                                                                                                                                    &                           & Arm joints: shoulders/elbow/forearm twist Hand/wrist/fingers                     & 8                       &                                        & Optical 3D                                                 & RMSE                                                                                &                                          \\ 
\hdashline
Research on Human Motion Monitoring Method Based on Multi-Joint Constraint Filter Model~\cite{Yang2021}                                                                                                                      &                           &                                                                                  & 1                       &                                        & Goniometer                                                 & RMSE                                                                                & Max. error                                \\ 
\hdashline
Rhythmic EKF for pose estimation during gait~\cite{Joukov2015}                                                                                                                                                               & Hip/knee/ankle            &                                                                                  & 3                       &                                        & Optical 3D                                                 & RMSE                                                                                &                                          \\ 
\hdashline
Rhythmic Extended Kalman Filter for Gait Rehabilitation Motion Estimation and Segmentation~\cite{Joukov2018}                                                                                                                 & Hip/knee/ankle            &                                                                                  & 5                       & 2 post-stroke                          & Optical 3D                                                 & RMSE                                                                                &                                          \\ 
\hdashline
Rigid body motion capturing by means of a wearable inertial and magnetic MEMS sensor assembly-from reconstitution of the posture toward dead reckoning: An application in bio-logging~\cite{Fourati2017}                     & Feet                      &                                                                                  & 1                       &                                        & Visual 2D                                                  & MAE                                                                                 &                                          \\ 
\hdashline
Robust and Accurate Capture of Human Joint Pose Using an Inertial Sensor~\cite{Pathirana2018}                                                                                                                                &                           & Head/neck/scapula Back/trunk/torso                                               & 14                      &                                        & Optical 3D \& RGB + depth                                   & RMSE                                                                                &                                          \\ 
\hdashline
Sensorial system for obtaining the angles of the human movement in the coronal and sagittal anatomical planes~\cite{Alvarado2017}                                                                                            &                           & Back/trunk/torso Arm                                                             & 1                       &                                        & Other                                              & RMSE                                                                                &                                          \\ 
\hdashline
Shoulder and elbow joint angle estimation for upper limb rehabilitation tasks using low-cost inertial and optical sensors~\cite{Alizadegan2017}                                                                              &                           & Arm joints: shoulders/elbow/forearm twist                                        & 1                       &                                        & Goniometer                                                 & RMSE                                                                                &                                          \\ 
\hdashline
Shoulder and elbow joint angle tracking with inertial sensors~\cite{ElGohary2012}                                                                                                                                            &                           & Arm joints: shoulders/elbow/forearm twist                                        & 8                       &                                        & Optical 3D                                                 & RMSE                                                                                &                                          \\ 
\hdashline
Synergy-based knee angle estimation using kinematics of thigh~\cite{Liang2021}                                                                                                                                               & Hip/knee/ankle            &                                                                                  & 8                       &                                        & IMU                                  & RMSE                                                                                &                                          \\ 
\hdashline
System for measurement of joint range of motion using inertial sensors~\cite{Sharma2017}                                                                                                                                     & Hip/knee/ankle            &                                                                                  & 1                       &                                        & Goniometer                                                 & MAE                                                                       &                                          \\ 
\hdashline
The head mouse - Head gaze estimation 'In-the-Wild' with low-cost inertial sensors for BMI use~\cite{Sim2013}                                                                                                                &                           & Head/neck/scapula                                                                & 1                       &                                        & IMU                                  & RMSE                                                                                &                                          \\ 
\hdashline
The manumeter: A wearable device for monitoring daily use of the wrist and fingers~\cite{Friedman2014}                                                                                                                       &                           & Hand/wrist/fingers                                                               & 7                       &                                        & Goniometer                                                 & MAE                                                                         &                                          \\ 
\hdashline
The online estimation of the joint angle based on the gravity acceleration using the accelerometer and gyroscope in the wireless networks~\cite{Ding2020}                                                                    & Hip/knee/ankle            &                                                                                  & 1                       &                                        & Optical 3D                                                 & RMSE                                                                                &                                          \\ 
\hdashline
The use of synthetic IMU signals in the training of deep learning models significantly improves the accuracy of joint kinematic predictions~\cite{SharifiRenani2021}                                                         & Hip/knee/ankle            &                                                                                  & 30                      & 17 knee arthroplasty                   & Optical 3D                                                 & RMSE                                                                                &                                          \\ 
\hdashline
Three dimensional gait analysis using wearable acceleration and gyro sensors based on quaternion calculations~\cite{Tadano2013}                                                                                              & Hip/knee/ankle            &                                                                                  & 5                       &                                        & Optical 3D                                                 & RMSE                                                                                &                                          \\ 
\hdashline
Time coherent full-body poses estimated using only five inertial sensors: Deep versus shallow learning~\cite{Wouda2019}                                                                                                      &                           &                                                                                  & 6                       &                                        & IMU                                  & MAE                                                                &                  \\ 
\hdashline
Towards a portable human gait analysis  monitoring system~\cite{Kumar2018}                                                                                                                                                   & Hip/knee/ankle            &                                                                                  & 3                       &                                        & Optical 3D                                                 & RMSE                                                                                &                                          \\ 
\hdashline
Towards kinematically constrained real time human pose estimation using sparse IMUs~\cite{Nagaraj2021}                                                                                                                       &                           &                                                                                  & 15                      &                                        & IMU                                  & MAE                                                                                 &                                          \\ 
\hdashline
Training Data Selection and Optimal Sensor Placement for Deep-Learning-Based Sparse Inertial Sensor Human Posture Reconstruction~\cite{Zheng2021}                                                                            &                           &                                                                                  & 4                       &                                        & IMU                                  & MAE                                                        &                                          \\ 
\hdashline
Ubiquitous human upper-limb motion estimation using wearable sensors~\cite{Zhang2011b}                                                                                                                                       &                           & Arm joints: shoulders/elbow/forearm twist                                        & 4                       &                                        & Optical 3D                                                 & RMSE                                                                                &                                          \\ 
\hdashline
Unrestrained measurement of arm motion based on a wearable wireless sensor network~\cite{Lee2010}                                                                                                                            &                           & Arm joints: shoulders/elbow/forearm twist                                        & 1                       &                                        & Goniometer                                                 & MAE                        &                                          \\ 
\hdashline
Upper limb joint angle measurement in occupational health~\cite{Alvarez2016}                                                                                                                                                 &                           & Arm joints: shoulders/elbow/forearm twist                                        & 1                       &                                        & Visual 2D                                                  & Typical deviation                                                                   & MAE                               \\ 
\hdashline
Upper limb joint angle tracking with inertial sensors~\cite{ElGohary2011}                                                                                                                                                    &                           & Arm joints: shoulders/elbow/forearm twist                                        & 1                       &                                        & Optical 3D                                                 & Corr. coeff.                                                                         &                                          \\ 
\hdashline
Use of multiple wearable inertial sensors in upper limb motion tracking~\cite{Zhou2008}                                                                                                                                      &                           & Arm joints: shoulders/elbow/forearm twist Hand/wrist/fingers                     & 4                       &                                        & Optical 3D                                                 & RMSE                                                                                &                                          \\ 
\hdashline
Validation of a low-cost inertial exercise tracker~\cite{Salehi2020}                                                                                                                                                         & Leg Hip/knee/ankle        &                                                                                  & 7                       &                                        & Optical 3D                                                 & RMSE                                                                                &                                          \\ 
\hdashline
Visual and quantitative analysis of lower limb 3D gait posture using accelerometers and magnetometers~\cite{Liu2010b}                                                                                                        & Hip/knee/ankle            &                                                                                  & 5                       &                                        & Optical 3D                                                 & RMSE                                                                                & Corr. coeff.                   \\ 
\hdashline
Visual estimation of lower limb motion using physical and virtual sensors~\cite{Liu2010c}                                                                                                                                    & Hip/knee/ankle            &                                                                                  & 3                       &                                        & Optical 3D                                                 & RMSE                                                                                & Corr. coeff.                   \\ 
\hdashline
Wearable inertial sensor system towards daily human kinematic gait analysis: Benchmarking analysis to MVN BIOMECH~\cite{Figueiredo2020}                                                                                      & Hip/knee/ankle            &                                                                                  & 11                      &                                        & IMU                                  & RMSE &                                                                                               \\
\hline
\end{longtable}
\end{landscape}
\clearpage
\twocolumn

\section*{Acknowledgment}
This work was supported by Junta de Comunidades de Castilla La Mancha (FrailAlert project, SBPLY/21/180501/000216), the
Spanish Ministry of Science and Innovation (INDRI project, PID2021-122642OB-C41) , Comunidad de Madrid (RACC project, CM/JIN/2021-016) and NEXTPERCEPTION European Union project funded by ECSEL Joint Undertaking (JU), under grant agreement No.876487 (ECSEL-2019-2-RIA).

\bibliography{references}
\bibliographystyle{ieeetr}

\end{document}